\documentclass[final]{siamltex}

\usepackage{amssymb,amsmath}
\usepackage[numbered]{mcode}
\usepackage{graphicx,url, booktabs, topcapt}
\usepackage{clrs+} 
\usepackage{algorithm}
\usepackage{subfig}
\usepackage[noend]{algpseudocode}
\usepackage{enumitem}
\setlist{noitemsep,leftmargin=*}
\usepackage{color}
\usepackage{multirow}

\graphicspath{{figures/}}
\usepackage{fixltx2e}
\MakeRobust{\Call}

\usepackage{listings}
\newcommand{\dimN}{n}
\newcommand{\dimM}{m}
\newcommand{\dimK}{k}
\newcommand{\dimL}{l}

\newcommand{\dnzc}{\id{nzc}}

\newcommand{\dnnz}{\id{nnz}}

\newcommand{\dth}{th}
\newcommand{\dlen}{\id{len}}

\newcommand{\AUX}{\mathsf{AUX}}

\newcommand{\qC}{\c{C}}

\newcommand{\matlab}{{\sc Matlab}}
\newcommand{\mA}{\mathbf{A}} 
\newcommand{\mB}{\mathbf{B}}
\newcommand{\mR}{\mathbf{R}}
\newcommand{\transpose}     {^{\mbox{\scriptsize \sf T}}}
\newcommand{\mBT}{\mathbf{B^T}}
\newcommand{\mRT}{\mathbf{R^T}}
\newcommand{\mC}{\mathbf{C}}
\newcommand{\flops}{\mathrm{flops}}
\newcommand{\rmat}{R-MAT}
\newcommand{\erdosrenyi}{Erd\H os-R\'{e}nyi}

\def\Cpp{C{}\texttt{++}~}
\newcommand{\SpGEMMThreeD}{Split-3D-SpGEMM}

\newcommand{\lilabel}[1]        {\label{li:#1}}
\newcommand{\liref}[1]      {line~\ref{li:#1}}
\newcommand{\Liref}[1]      {Line~\ref{li:#1}}

\newcommand{\vsf}[0]{\vspace{-9pt}} 

\title{Exploiting Multiple Levels of Parallelism in Sparse Matrix-Matrix Multiplication}

\author{Ariful Azad\thanks{CRD, Lawrence Berkeley National Laboratory, CA ({\tt azad@lbl.gov}).}
\and Grey Ballard\thanks{Sandia National Laboratories, Livermore, California.  Current address: Computer Science Department, Wake Forest University, Winston Salem, North Carolina  ({\tt ballard@wfu.edu}).}
  \and Aydin Bulu\qC\thanks{CRD, Lawrence Berkeley National Laboratory, CA ({\tt abuluc@lbl.gov}).}
  \and James Demmel\thanks{EECS, University of California, Berkeley, CA ({\tt demmel@eecs.berkeley.edu}).}
  \and Laura Grigori\thanks{INRIA Paris-Rocquencourt, Alpines, France ({\tt laura.grigori@inria.fr}).}
  \and Oded Schwartz\thanks{The Hebrew University, Israel ({\tt odedsc@cs.huji.ac.il}). }
    \and Sivan Toledo\thanks{Blavatnik School of Computer Science, Tel Aviv University, Israel ({\tt stoledo@tau.ac.il}). }
  \and Samuel Williams\thanks{CRD, Lawrence Berkeley National Laboratory, CA ({\tt swwilliams@lbl.gov}). }}

\begin{document}

\maketitle

\begin{abstract}
Sparse matrix-matrix multiplication (or SpGEMM) is a
key primitive for many high-performance graph algorithms as well 
as for some linear solvers, such as algebraic multigrid. 
The scaling of existing parallel implementations of SpGEMM is heavily
bound by communication. Even though 3D (or 2.5D) algorithms have been 
proposed and theoretically analyzed in the flat MPI model on \erdosrenyi\ matrices, those algorithms
had not been implemented in practice and their complexities had not been analyzed
for the general case. In this work, we present the first implementation of the 3D 
SpGEMM formulation that exploits multiple (intra-node and inter-node) levels of parallelism, achieving 
significant speedups over the state-of-the-art publicly available codes at all levels of concurrencies. 
We extensively evaluate our implementation and identify bottlenecks that should be subject to further research.
\end{abstract}



\pagestyle{myheadings}
\thispagestyle{plain}
\markboth{AZAD ET AL.}{MULTIPLE LEVELS OF PARALLELISM IN SPGEMM}

\section{Introduction}
Multiplication of two sparse matrices (SpGEMM) is a key operation for high-performance graph computations in the language
of linear algebra~\cite{GALLA, mattson2013standards}. Examples include
graph contraction~\cite{unifiedstarp}, betweenness centrality~\cite{combblas}, 
Markov clustering~\cite{vandongen00}, peer pressure clustering~\cite{shahthesis}, triangle counting~\cite{trianglegabb15},
and cycle detection~\cite{cycle}. SpGEMM is also used in scientific computing. For instance, it
is often a performance bottleneck for Algebraic Multigrid (AMG), where it is used in the set-up phase 
for restricting and interpolating matrices~\cite{bell2012exposing}. Schur complement methods in hybrid linear 
solvers~\cite{YamazakiSchurComp10} also require fast SpGEMM. In electronic structure calculations, 
linear-scaling methods exploit Kohn's ``nearsightedness'' principle of electrons in many-atom systems~\cite{kohn1996density}. 
SpGEMM and its approximate versions are often the workhorse of these computations~\cite{bock2013optimized,borvstnik2014sparse}. 

We describe new parallel implementations of the SpGEMM
kernel, by exploiting multiple levels of parallelism. 
We provide the first complete implementation and large-scale results of a ``3D algorithm'' that
asymptotically reduces communication costs compared to the state-of-the-art 2D algorithms.
The name ``3D'' derives from the parallelization across all 3 dimensions of the iteration space.
While 2D algorithms like Sparse SUMMA \cite{gemmexp} are based on a 2D decomposition of the output matrix with computation following an ``owner computes'' rule, 
a 3D algorithm also parallelizes the computation of individual output matrix entries. 
Our 3D formulation relies on splitting (as opposed to replicating) input submatrices across processor layers.

While previous work~\cite{spaa13} analyzed the 
communication costs of a large family of parallel SpGEMM algorithms and provided lower-bounds on random matrices, it did 
not present any experimental results. In particular, the following questions were left unanswered:

\begin{itemize}
\item What is the effect of different communication patterns on relative scalability of these algorithms? The analysis 
was performed in terms of ``the number of words moved per processor'', which did not take into account important factors such 
as network contention, use of collectives, the relative sizes of the communicators, etc.
\item What is the effect of in-node multithreading? By intuition, one can expect a positive effect due to reduced network contention 
and automatic data aggregation as a result of  in-node multithreading, but those have not been evaluated before.
\item What is the role of local data structures and local algorithms? In particular, what is the right data structure to store
local sparse matrices in order to multiply them fast using a single thread and multiple threads? How do we merge local triples 
efficiently during the reduction phases?
\item How do the algorithms perform on real-world matrices, such as those with skewed degree distributions?
\end{itemize}

This paper addresses these questions by presenting the first implementation of the 3D SpGEMM 
formulation that exploits both the additional third processor grid dimension and the in-node multithreading aspect.
In particular, we show that the third processor grid dimension navigates a tradeoff between communication of the \emph{input} matrices and communication of 
the \emph{output} matrix. We also show that in-node multithreading, with efficient shared-memory parallel kernels, can significantly enhance scalability.
In terms of local data structures and algorithms, we use a priority queue to 
merge sparse vectors for in-node multithreading. This eliminates thread scaling bottlenecks which were due to 
asymptotically increased working set size as well as the need to modify the data structures for cache efficiency. 
To answer the last question, we benchmark our algorithms on real-world matrices coming from a variety of applications.
Our extensive evaluation via large-scale experiments exposes bottlenecks and provides new avenues for research. 

Section~\ref{sec:spgemm} summarizes earlier results on various parallel SpGEMM formulations. 
Section~\ref{sec:parspgemm} presents the distributed-memory algorithms implemented for this work, 
as well as the local data structures and operations in detail.
In particular, our new 3D algorithm, \SpGEMMThreeD{}, is presented in Section~\ref{sec:25dspgemm}.
Section~\ref{sec:expparallel} gives an extensive performance evaluation of these implementations using 
large scale parallel experiments, including a performance comparison with similar primitives offered by other publicly available 
libraries such as Trilinos and Intel Math Kernel Library (MKL). 
Various implementation decisions and their effects on performance are also detailed.

\section{Notation}
Let $\mA \in \mathbb{S}^{\dimM \times \dimK}$ be a sparse rectangular matrix of elements from a semiring $\mathbb{S}$. 
We use $\dnnz(\mA)$ to denote the number of nonzero elements in $\mA$. When the
matrix is clear from context, we drop the parenthesis and simply use $\dnnz$. For sparse matrix indexing, we use the convenient 
${\textrm \matlab }$ colon notation, where $\mA(:,i)$ denotes the $i$\dth\ column, $\mA(i,:)$ denotes  the $i$\dth\ row, and $\mA(i,j)$ denotes the 
element at the $(i,j)$\dth\ position of matrix $\mA$. 
Array and vector indices are 1-based throughout this paper. 
The length of an array $\mathsf{I}$,  denoted by $\dlen(\mathsf{I})$, is equal to its number of elements.
For one-dimensional arrays, $\mathsf{I}(i)$ denotes the $i$\dth\ component of the array.
$\mathsf{I}(j:k)$ defines the range $\mathsf{I}(j), \mathsf{I}(j+1),\ldots,\mathsf{I}(k)$ and is also applicable to matrices.

We use $\flops(\mA, \mB)$, pronounced ``flops'', 
to denote the number of nonzero arithmetic operations required when computing the product of matrices $\mA$ and $\mB$. 
When the operation and the operands are clear from context, we simply use $\flops$. We acknowledge that semiring operations
do not have to be on floating-point numbers (e.g. they can be on integers or Booleans) but we nevertheless use $\flops$ as opposed
to $\mathrm{ops}$ to be consistent with existing literature. 

In our analysis of parallel running
time, the latency of sending a message over the communication interconnect is $\alpha $, 
and the inverse bandwidth is $\beta $, both expressed as multiples of the time for a floating-point 
operation (also accounting for the cost of cache misses and memory indirections associated with that floating point operation).
Notation $f(x) = \Theta(g(x))$ means that $f$ is bounded asymptotically by $g$ both above and below.
We index a 3D process grid with $P(i,j,k)$. Each 2D slice of this grid $P(:,:,k)$ with the third dimension fixed is called a 
process ``layer'' and each 1D slice of this grid $P(i,j,:)$ with the first two dimensions fixed is called a process ``fiber''.

\section{Background and Related Work}
\label{sec:spgemm}

The classical serial SpGEMM algorithm for general sparse matrices was first described by Gustavson~\cite{gust:78}, 
and was subsequently used in Matlab~\cite{smatlab} and CSparse~\cite{davisbook}. 
For computing the product $\mC = \mA \mB $, where $\mA \in \mathbb{S}^{\dimM \times \dimL}$,
$\mB \in \mathbb{S}^{\dimL \times \dimN}$ and $\mC \in \mathbb{S}^{\dimM \times \dimN}$, 
Gustavson's algorithm runs in $O(\flops + \dnnz+\dimM + \dimN)$ time, which is optimal when $\flops$ is larger than $\dnnz$, $\dimM$,
and $\dimN$. It uses the popular 
compressed sparse column (CSC) format for representing its sparse matrices. Algorithm~\ref{alg:cscgemm} gives the pseudocode for this column-wise serial algorithm for SpGEMM. 

\begin{algorithm}
\begin{algorithmic}[1]
\Procedure{Columnwise-SpGEMM}{$\mA, \mB, \mC$}
\For{$k \gets 1$ to $\dimN$}
	\For{$j$ where $\mB(j,k) \neq 0 $}	
	\State $ \mC(:, k) \gets \mC(:, k) + \mA(:, j) \cdot \mB(j, k) $
	\EndFor
\EndFor
\EndProcedure
\end{algorithmic}
\caption{Column-wise formulation of serial matrix multiplication} \label{alg:cscgemm}
\end{algorithm}

McCourt et al.~\cite{mccourt2015sparse} target $\mA \mBT$ and $\mathbf{R} \mA \mRT$ operations in the specific context of Algebraic Multigrid (AMG). A coloring of the output matrix $\mC$ finds {\em structurally orthogonal} columns that can be computed simultaneously. 
Two columns are structurally orthogonal if the inner product of their structures (to avoid numerical cancellation) is zero.
They use matrix colorings to restructure $\mBT$ and $\mRT$ into
dense matrices by merging non-overlapping sparse columns that do not contribute to the same nonzero in the output matrix. 
They show that this approach would incur less memory traffic than performing sparse inner products by
a factor of $n/n_{color}$ where $n_{color}$ is the number of colors used for matrix coloring. 
However, they do not analyze the memory traffic of other formulations
of SpGEMM, which are known to outperform sparse inner products~\cite{sparsechapter}. 
In particular, a column-wise formulation of SpGEMM using CSC incurs only $O(\dnnz / L + \flops)$ cache misses where $L$ is the size of the cache line. Consider the
matrix representing the \erdosrenyi\ graph $G(n,p)$, where each edge (nonzero) in the graph (matrix) is present with probability $p$ independently from each other. For $p=d/n$ where $d \ll n$, in expectation $\dimN d$ nonzeros are uniformly distributed in an $\dimN$-by-$\dimN$ sparse matrix. 
In that case, SpGEMM does $O(d^2 \dimN)$ cache misses compared to the $O(d \, \dimN \, n_{color})$ cache misses of the algorithm by McCourt et al. Hence, the column-wise approach not only bypasses the need for coloring, it also performs better for $d \leq n_{color}$, which is a common case. Furthermore, their method requires precomputing the nonzero structure of the output matrix, which is asymptotically as hard as computing SpGEMM without coloring in the first place.

There has been a flurry of activity in developing algorithms and implementations of SpGEMM for Graphics Processing Units (GPUs). Among those,
the algorithm of Gremse et al.~\cite{gremse2015gpu} uses the row-wise formulation of SpGEMM. By contrast, Dalton et al.~\cite{Dalton:2015:OSM}
uses the data-parallel ESC (expansion, sorting, and contraction) formulation, which is based on outer products. One downside
of the ESC formulation is that expansion might create $O(\flops)$ intermediate storage in the worst case, depending on the number of
additions performed immediately in shared memory when possible, which might be asymptotically larger than 
the sizes of the inputs and outputs combined. The recent work of Liu and Vinter is currently the fastest implementation on GPUs and it 
also addresses heterogenous CPU-GPU processors~\cite{Liu2015}.

In distributed memory, under many definitions of scalability, all known parallel SpGEMM algorithms are unscalable due to increased communication costs relative to arithmetic operations. 
For instance, there is no way to keep the parallel efficiency ($\mathit{PE}$) fixed for any constant $1 \geq \mathit{PE} > 0$ as we increase the number of 
processors~\cite{kumar1994analyzing}.  Recently, two attempts have been made to model the communication costs of SpGEMM in a more fine grained manner. 
Akbudak and Aykanat~\cite{akbudak2014simultaneous} proposed the first hypergraph model for outer-product formulation of SpGEMM. 
Unfortunately, a symbolic SpGEMM computation has to be performed initially as the hypergraph 
model needs full access to the computational pattern that forms the output matrix.
Ballard et al.~\cite{BallardDKS15} recently proposed hypergraph models 
for a class of SpGEMM algorithms more general than Akbudak and Aykanat considered. 
Their model also requires the sparsity structure of the output matrix and the number 
of vertices in the hypergraph is $O(\flops)$, making the approach impractical. 

In terms of in-node parallelism via multithreading, there has been relatively little work. Gustavson's algorithm is not thread scalable because its intermediate
working set size is $O(\dimN)$ per thread, requiring a total of $O(\dimN  t)$ intermediate storage, which can be larger than the matrices themselves for high thread
counts. This intermediate data structure is called the sparse accumulator (SPA)~\cite{smatlab}. Nevertheless, it is possible to get good performance 
out of a multithreaded parallelization of Gustavson's algorithm in current platforms, provided that accesses to SPA are further ``blocked'' for matrices with large dimensions, in order to decrease cache miss rates. In a recent work, this is achieved by partitioning the data structure of the second matrix $\mB$ by 
columns~\cite{patwary2015parallel}. 

We also mention that there has been significant research devoted to dense matrix multiplication in distributed-memory settings.
In particular, the development of so-called 3D algorithms for dense matrix multiplication spans multiple decades; see \cite{DNS81,ITT04,SPvdG13,SD11} and the references therein.
Many aspects of our 3D algorithm for sparse matrix multiplication are derived from the dense case, though there are important differences as we detail below.

\section{Distributed-memory SpGEMM}
\label{sec:parspgemm}
\begin{figure}[ht]
  \centering
     \subfloat[][1D]{\includegraphics[scale=.25]{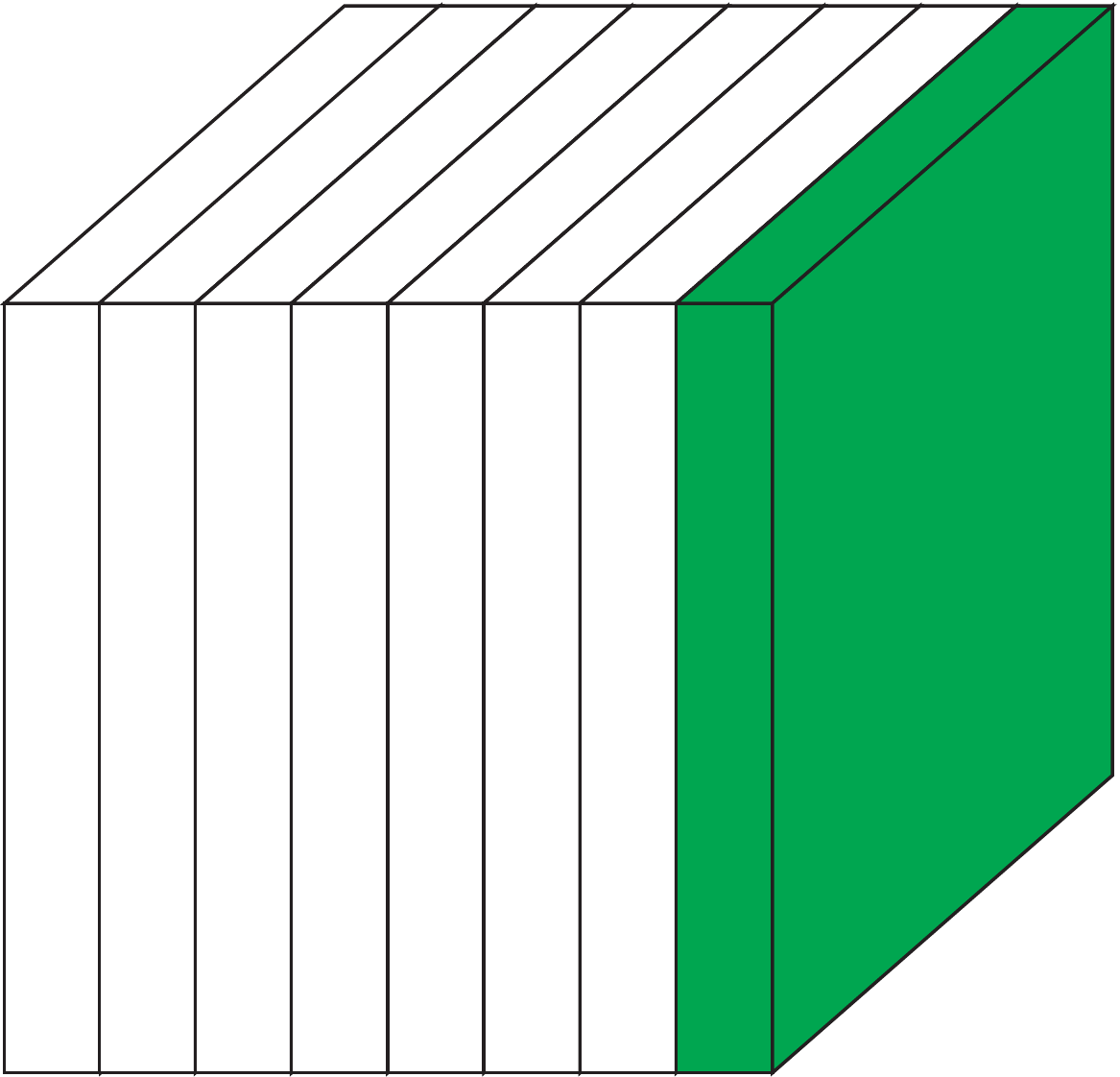}\label{fig:1d}} \qquad
          \subfloat[][2D]{\includegraphics[scale=.25]{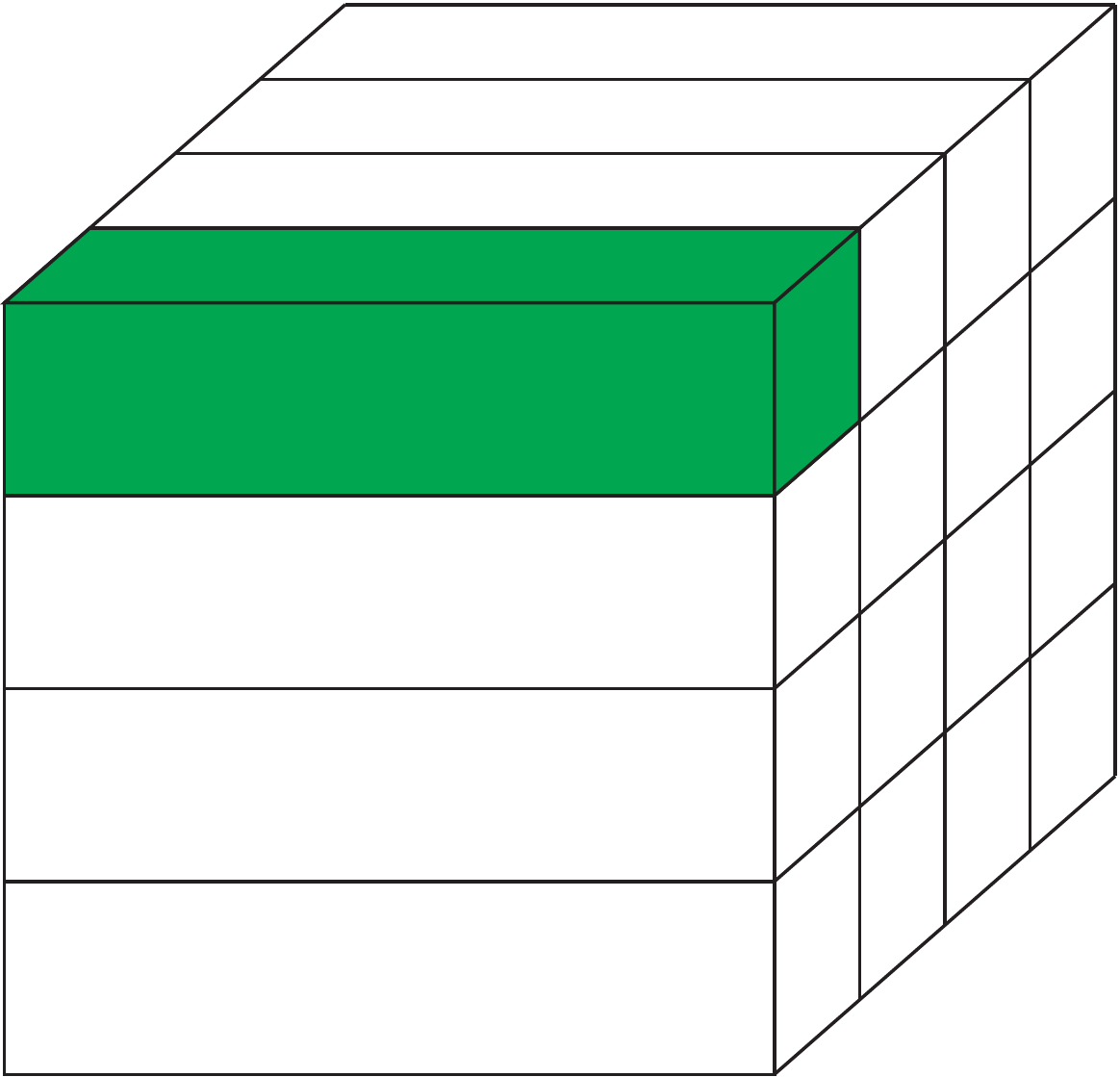}\label{fig:2d}} \qquad
     \subfloat[][3D]{\includegraphics[scale=.25]{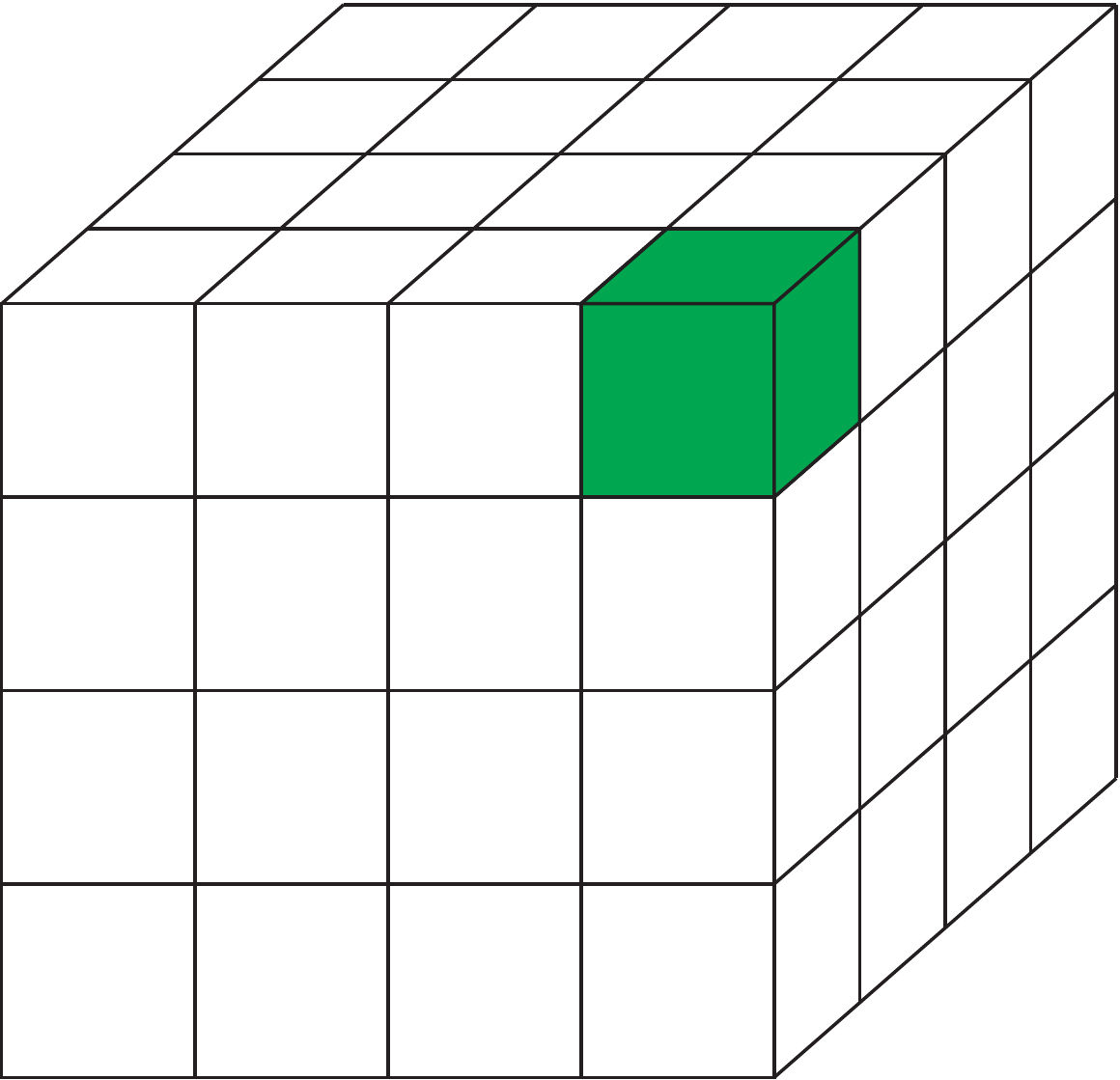}\label{fig:3d}}
       \caption{Partitioning the work cube to processes. Image reproduced for clarity~\cite{spaa13}.} 
       \label{fig:distributions}
\end{figure}

We categorize algorithms based on how they partition ``work'' (scalar multiplications) among processes, as we first advocated recently~\cite{spaa13}.
The work required by SpGEMM can be conceptualized by a cube that is sparsely populated by ``voxels'' that correspond to nonzero scalar multiplications. 
The algorithmic categorization is based on how these voxels are assigned to processes, which is illustrated in Figure~\ref{fig:distributions}.
1D algorithms assign a block of $n$-by-$n$-by-$1$ ``layers'' of this cube to processes. 
In practice, this is realized by having each process store a block of rows or columns 
of an $\dimM$-by-$\dimN$ sparse matrix, though the 1D/2D/3D categorization is separate from the data distribution. 
With correctly chosen data distributions, 1D algorithms communicate entries of only one of the three matrices.

2D algorithms assign a set of $1$-by-$1$-by-$n$ ``fibers'' of this cube to processes. 
In many practical realizations of 2D algorithms, processes are logically organized as a rectangular $p = p_r \times p_c$ process grid, 
so that a typical process is named $P(i,j)$. 
Submatrices are assigned to processes according to a 2D block decomposition:
For a matrix $\mathbf{M}  \in \mathbb{S}^{\dimM \times \dimN}$, processor $P(i,j)$ stores the submatrix $\mathbf{M}_{ij}$ of dimensions $(m/p_r)\times (n/p_c)$ in its local memory. 
With consistent data distributions, 2D algorithms communicate entries of two of the three matrices.

3D algorithms assign subcubes (with all 3 dimensions shorter than $\dimN$) to processes, which 
are typically organized on a $p = p_r \times p_c \times p_l$ grid and indexed by $P(i,j,k)$.
3D algorithms communicate entries of $\mA$ and $\mB$, as well as the (partial sums of the) intermediate products of $\mC$.
While many ways of assigning submatrices to processes on a 3D process grid exist, including replicating
each $\mA_{ij}$ along the process fiber $P(i,j,:)$, our work focuses on a memory-friendly split decomposition.
In this formulation, $P(i,j,k)$ owns the following $\dimM/p_r \times \dimN/(p_c p_l)$ submatrix of $\mathbf{A}  \in \mathbb{S}^{\dimM \times \dimN}$:
$$\mathbf{A}(i \dimM/p_r: (i+1) \dimM/p_r-1, j \dimN/p_c+k \dimN/(p_c p_l) : j \dimN/p_c + (k+1) \dimN/(p_c p_l)-1).$$
The distribution of matrices $\mB$ and $\mC$ are analogous. This distribution is memory friendly because it does not replicate 
input or output matrix entries, which is in contrast to many 3D algorithms where the input or the output is explicitly replicated.


\subsection{Sparse SUMMA Algorithm}
\label{sec:sparsesumma}
We briefly remind the reader of the Sparse SUMMA algorithm~\cite{icpp08} for completeness as it will form the base of our 3D discussion. 
Sparse SUMMA is based on one formulation of the dense SUMMA algorithm~\cite{summa}. The processes are logically organized on a
$p_r \times p_c$ process grid. The algorithm proceeds in stages where each stage involves the broadcasting of $n/p_r \times b$ submatrices of $\mA$ by their 
owners along their process row, and the broadcasting of $b \times n/p_c$ submatrices of $\mB$ by their owners along their process column.
The recipients multiply the submatrices they received to perform a rank-$b$ update on their piece of the output matrix $\mC$. 
The rank-$b$ update takes the form of a \emph{merge} in the case of sparse matrices; several rank-$b$ updates can be done together using a multiway 
merge as described in Section \ref{sec:multiway}.
We will refer to this as a ``SUMMA stage'' for the rest of the paper.
Here, $b$ is a blocking parameter, which can be as large as the inner submatrix dimension. A more complete description of the algorithm and its
general implementation for rectangular matrices and process grids can be found in an earlier work~\cite{gemmexp}.

\subsection{In-node Multithreaded SpGEMM Algorithm}
\label{sec:heapspgemm}

Our previous work~\cite{ipdps08} shows that the standard compressed sparse column or row 
(CSC or CSR) data structures are too wasteful for storing the local submatrices arising from a 2D decomposition.
This is because the local submatrices are {\em hypersparse}, 
meaning that the ratio of nonzeros to dimension asymptotically approaches zero as the number of processors increase.
The total memory across all processors for CSC format would be $O(\dimN\sqrt{p}+\dnnz)$, 
as opposed to $O(\dimN+\dnnz)$ memory to store the whole matrix in CSC on a single processor. 

This observation applies to 3D algorithms as well because their execution is reminiscent of running
a 2D algorithm on each processor layer $P(:,:,k)$. Thus, local data structures used within 2D and 3D algorithms must respect hypersparsity.

Similarly, any algorithm whose complexity depends on matrix dimension, such as Gustavson's serial SpGEMM algorithm, 
is asymptotically too wasteful to be used as a computational kernel for multiplying the hypersparse submatrices.
We use \proc{HeapSpGEMM}, first presented as Algorithm 2 of our earlier work~\cite{ipdps08}, which operates on the strictly $O(\dnnz)$ 
{\em doubly compressed sparse column (DCSC)} data structure, 
and its time complexity does not depend on the matrix dimension. 
DCSC~\cite{ipdps08} is a further compressed version of CSC where repetitions in the column pointers array, 
which arise from empty columns, are not allowed. Only columns that have at least one nonzero are represented, together with their column indices.  
DCSC is essentially a sparse array of sparse columns, whereas CSC is a dense array of sparse columns.
Although not part of the essential data structure, DCSC can support fast column indexing by building 
an $\AUX$ array that contains pointers to nonzero columns (columns that have at least one nonzero element) in linear time.

\begin{figure}[t]
\centering  
\includegraphics[scale=0.45]{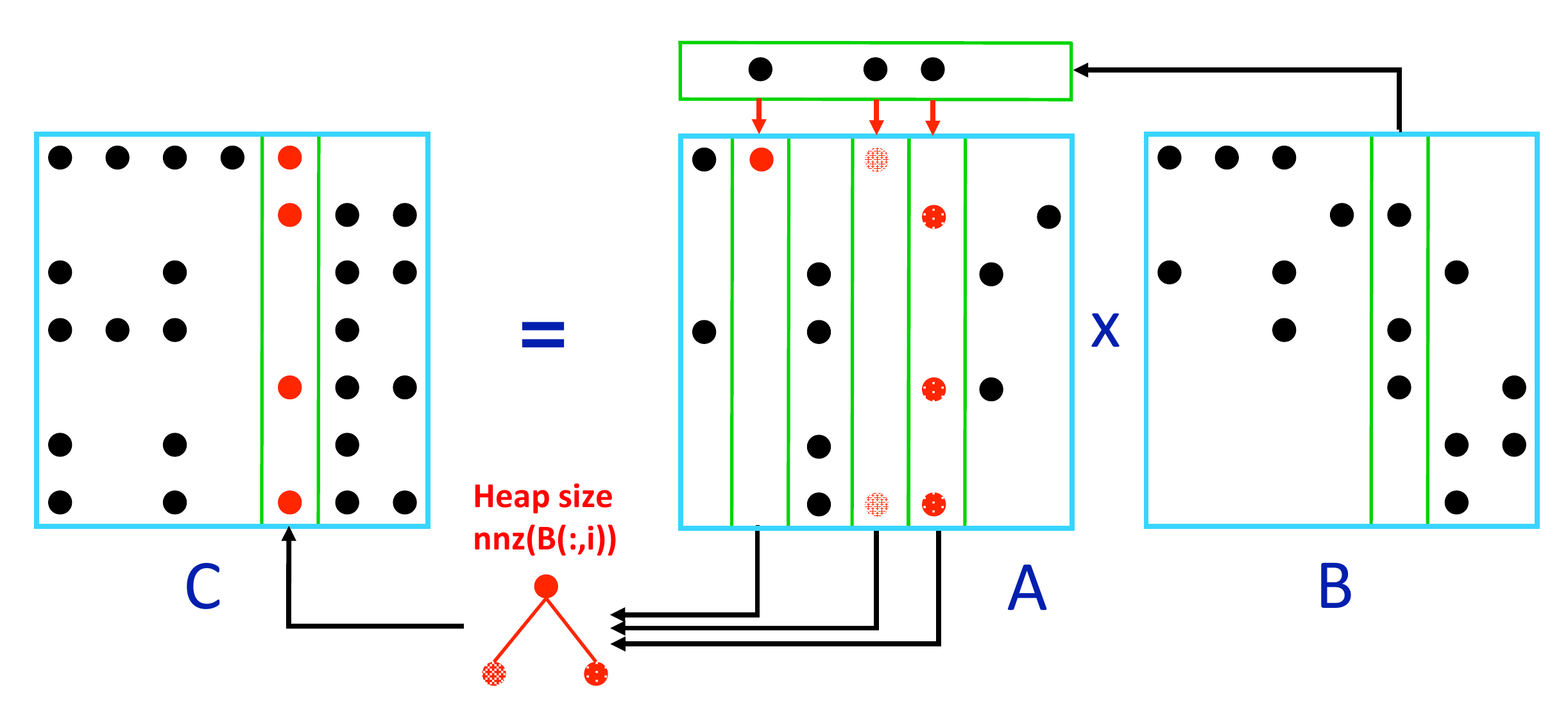}  
\caption[Multiply sparse matrices column-by-column using a priority queue]
{Multiplication of sparse matrices stored by columns~\cite{ipdps08}. Columns of $\mA$ are accumulated as specified by the non-zero entries in a column of $\mB$ using a priority queue (heap) indexed by the row indices.  
The contents of the heap are stored into a column of $\mC$ once all required columns are accumulated. } 
\label{fig:cscmultpic}
\end{figure}
 
Our \proc{HeapSpGEMM} uses a heap-assisted column-by-column formulation whose time complexity is
$$\sum_{j=0}^{\dnzc(B)} O\bigl(\flops(\mC(:,j)) \, \log{\dnnz(\mB(:,j))\bigr)},$$

\noindent
where $\dnzc(\mB)$ is the number of columns of $\mB$ that are not entirely zero, 
$\flops(\mC(:,j))$ is the number of nonzero multiplications and additions required to generate the $j$th column of $\mC$. The execution of this algorithm is 
illustrated in Figure~\ref{fig:cscmultpic}, which differs from Gustavson's formulation in its use of a heap (priority queue) as opposed to a sparse accumulator (SPA). 

Our formulation is more suitable for multithreaded execution where we parallelize over the columns of $\mC$ and 
each thread computes $\mA$ times a subset of the columns of $\mB$.
SPA is an $O(\dimN)$ data structure, hence a multithreaded parallelization over columns of $\mC$ of the SPA-based algorithm would require $O(\dimN t) $ space for $t$ threads. 
By contrast, since each heap in \proc{HeapSpGEMM} is of size $O(\dnnz(\mB(:,j))$,
the total temporary memory requirements of our multithreaded algorithm are always strictly smaller than the space required to hold one of the inputs, namely $\mB$. 

\subsection{Multithreaded Multiway Merging and Reduction}
\label{sec:multiway}
Each stage of Sparse SUMMA generates partial result matrices that are summed together at the end of all stages to obtain the final result $\mC$.
In the 3D algorithm discussed in Section~\ref{sec:25dspgemm}, we also split $\mC$ across fibers of 3D grid, and the split submatrices are summed together by each process on the fiber. 
To efficiently perform these two summations, we represent the intermediate matrices as lists of triples, where each triple $(i, j, val)$ stores the row index, column index, and value of a nonzero entry, respectively.
Each list of triples is kept sorted lexicographically by the $(j,i)$ pair so that the $j$th column comes before the $(j{+}1)$st column. 
We then perform the summation of sparse matrices by merging the lists of triples that represent the matrices.
The merging also covers the reduction of triples with repeated indices. 

To perform a $k$-way merge on $k$ lists of triples $T_1, T_2, ...,T_k$, we maintain a heap of size $k$ that stores the current lexicographically minimum entry, based on $(j,i)$ pairs, in each list of triples. 
In addition to keeping triples, the heap also stores the index of the source list from where each triple was inserted into the heap.
The multiway merge routine finds the minimum triple $(i^*, j^*, val^*)$ from the heap and merges it into the result.
When the previous triple in the merged list has the same pair of indices $(i^*,j^*)$, the algorithm simply adds the values of these two triples, reducing the index-value pairs with repeated indices.
If $(i^*, j^*, val^*)$ is originated from $T_l$, the next triple from $T_l$ is inserted into the heap.
Hence, the time complexity of a $k$-way merge is
$$\sum_{l=1}^{k} O\bigl( \dnnz(T_l) \log{k} \bigr),$$

\noindent
where $\dnnz(T_l)$ is the number of nonzero entries in $T_l$.
When multithreading is employed, each thread merges a subset of columns from $k$ lists of triples using the same $k$-way merge procedure described earlier.
If a thread is responsible for columns $j_p$ to $j_q$, these column indices are identified from each list via a binary search.
For better load balance, we ensure there is enough parallel slackness~\cite{valiant1989optimally} by splitting
 the lists into more parts than the number of threads and merging the columns in each part by a single thread. 
 In our experiments, we created $4 t$ parts when using $t$ threads and employed dynamic thread scheduling.

\subsection{\SpGEMMThreeD{} Algorithm}
\label{sec:25dspgemm}
\begin{figure}
\begin{center}
\includegraphics[scale=0.55]{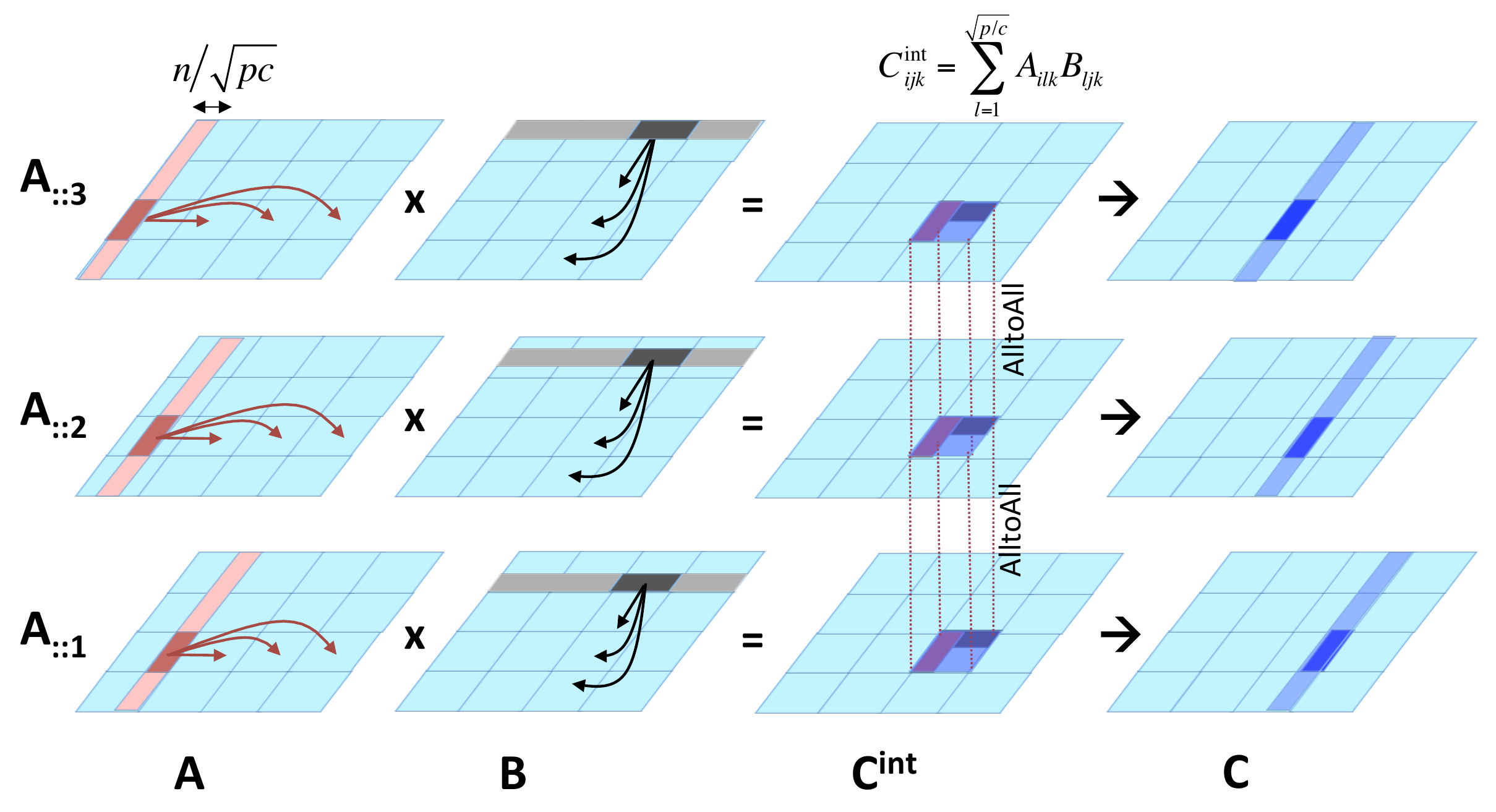}
\end{center}
\caption{Execution of the \SpGEMMThreeD{} algorithm for sparse matrix-matrix multiplication $\mC = \mA \cdot \mB$ on a $\sqrt{p/c} \times \sqrt{p/c} \times c$ process grid. 
Matrices  $\mA$, $\mB$, and $\mC^{int}$ matrices are shown during the first stage of the algorithm execution 
(the broadcast and the local update, i.e.\, one ``SUMMA stage''). The transition from $\mC^{int}$ to $\mC$
happens via an all-to-all followed by a local merge, after all $\sqrt{p/c}$ SUMMA stages are completed.
\label{fig:sp3D}}
\end{figure}

Our parallel algorithm is an iterative 3D algorithm that splits the submatrices along the third process grid dimension (of length $p_l$). 
This way, while there is no direct
relationship between the size of the third process dimension and the extra memory required for {\em the input}, 
the extra memory required by {\em the output}
is sparsity-structure dependent. If the output is sparse enough so that there are few or no intermediate products with repeated indices, 
then no extra memory is required. Recall that entries with repeated indices arise when more than one scalar multiplication 
$a_{ik} b_{kj}$ that results in a nonzero value contributes to the same output element $c_{ij}$.
The pseudocode of our algorithm, \proc{\SpGEMMThreeD}, is shown in Algorithm~\ref{alg:3D-SpGEMM} for the 
simplified case of $p_r = p_c =  \sqrt{p/c}$ and $p_l = c$.
The execution of the algorithm is illustrated in Figure~\ref{fig:sp3D}.

\begin{algorithm}
\begin{algorithmic}[1]
\Require $\mA \in \mathbb{S}^{\dimM \times \dimL},\mB \in \mathbb{S}^{\dimL \times \dimN}$: matrices on a  $\sqrt{p/c} \times \sqrt{p/c} \times c$ process grid
\Ensure $\mC \in \mathbb{S}^{\dimM \times \dimN}$: the product $\mA \mB$, similarly distributed. 
\Procedure{\SpGEMMThreeD{}}{$\mA, \mB, \mC$}
\State $ \id{locinndim} = \dimL/\sqrt{pc}$ \Comment{inner dimension of local submatrices} 
\For{all processes $P(i,j,k)$\  \InParallel} 
\State  $\hat \mB_{ijk}  \gets \Call{AlltoAll}{\mB_{ij:}, P(i,j,:)}$ \Comment{redistribution of $\mB$ across layers} \lilabel{Ball2all}
	\For{$r=1$ to $\sqrt{p/c}$} \Comment{$r$ is the broadcasting process column and row}
	\For{$q=1$ to $\id{locinndim}/b$} \Comment{$b$ evenly divides \id{locinndim}} 
	\State $\id{locindices} = (q-1) b: q b - 1$  \lilabel{loccol}
	\State	$\mA^{rem} \gets $ \Call{Broadcast}{$\mA_{irk}(:,\id{locindices}), P(i,:,k)$} 
	\lilabel{Abc}
	\State	$\mB^{rem} \gets $ \Call{Broadcast}{$\hat \mB_{rjk}(\id{locindices},:), P(:,j,k)$} 	\lilabel{Bbc}
	\State	$\mC_{ij:}^{int} \gets \mC_{ij:}^{int} + \Call{HeapSpGEMM}{\mA^{rem},\mB^{rem}}$ \lilabel{localmult}
	\EndFor	
	\EndFor
	\State	$\mC_{ijk}^{int} \gets \Call{AlltoAll}{\mC_{ij:}^{int}, P(i,j,:)}$	\lilabel{all2all}
	\State	$\mC_{ijk} \gets \Call{LocalMerge}{\mC_{ijk}^{int}}$	\lilabel{localmerge}
\EndFor
\EndProcedure
\end{algorithmic}
\caption{Operation $ \mC \gets \mA  \mB$ using \SpGEMMThreeD{}} \label{alg:3D-SpGEMM}
\end{algorithm}

The \proc{Broadcast}($\mA_{irk}, P(i,:,k)$) syntax means that the owner of $\mA_{irk}$ becomes the root and broadcasts
its submatrix to all the processes on the $i$th process row of the $k$th process layer. Similarly for \proc{Broadcast}($\hat \mB_{rjk}, P(:,j,k)$),
the owner of $\hat \mB_{rjk}$ broadcasts its submatrix to all the processes on the $j$th process column of the $k$th process layer. 
In \liref{loccol}, we find the local column (for $\mA$) and row (for $\hat \mB$) ranges for matrices that are 
to be broadcast during that iteration. They are significant only at the broadcasting processes, which can be determined 
implicitly from the first parameter of \proc{Broadcast}. In practice, we index $\hat \mB$ by columns as opposed to rows in order to 
obtain the best performance from
the column-based DCSC data structure. To achieve this, $\hat \mB$ gets locally transposed during redistribution in \liref{Ball2all}. 
Using DCSC, the expected cost of fetching 
$b$ consecutive columns of a matrix $\mA$ is $b$ plus the size (number of nonzeros) of the output.
Therefore, the algorithm asymptotically has the same computation cost for all blocking parameters $b$.

$\mC_{ij:}^{int}$ is the intermediate submatrix that contains nonzeros that can potentially belong to all the processes on the $(i,j)$th fiber 
$P(i,j,:)$. The \proc{AlltoAll} call in \liref{all2all} packs those nonzeros and sends them to their corresponding owners in the $(i,j)$th fiber. This results
in $\mC_{ijk}^{int}$ for each process $P(i,j,k)$, which contains only the nonzeros that belong to that process.
$\mC_{ijk}^{int}$ possibly contains repeated indices (i.e. multiple entries with the same index) that need to be merged and summed by the \proc{LocalMerge} call
in \liref{localmerge}, resulting in the final output. 

In contrast to dense matrix algorithms \cite{DNS81,ITT04,SPvdG13,SD11}, our sparse 3D formulation requires a more lenient trade-off between bandwidth-related communication costs
and memory requirements. As opposed to increasing the storage requirements by a factor of $p_l$, the relative cost of the 3D formulation is 
$\dnnz(\mC^{int}) / \dnnz(\mC)$, which is always upper bounded by $\flops(\mA, \mB) / \dnnz(\mC)$.

\subsection{Communication Analysis of the \SpGEMMThreeD{} Algorithm}
\label{sec:complexity}

For our complexity analysis, the previous work~\cite{spaa13} assumed that nonzeros of sparse $\dimN$-by-$\dimN$ input matrices are independently and identically distributed, with $d > 0$ nonzeros per row and column on the average. 
The sparsity parameter $d$ simplifies analysis by making different terms in the complexity comparable to each other. 
However, in order to capture the performance of more general matrix-matrix multiplication, we will analyze parallel complexity directly in terms of 
$\flops$ and the number of nonzeros in $\mA, \mB$, and $\mC$ without resorting to the sparsity parameter $d$.


Our algorithm can run on a wide range of configurations on a virtual 3D $p = p_r \times p_c \times p_l$ process grid.
To simplify the analysis, we again assume 
that each 2D layer of the 3D grid is square, i.e.\ $p_r=p_c$ and we use $c$ to denote the third dimension. 
Thus, we assume a $\sqrt{p/c} \times \sqrt{p/c} \times c$ process grid.


The communication in Algorithm~\ref{alg:3D-SpGEMM} consists of collective operations being performed on disjoint process fibers: simultaneous broadcasts in the first two process grid dimensions at \liref{Abc} and \liref{Bbc} and simultaneous all-to-alls in the third process grid dimension at \liref{Ball2all} and \liref{all2all}.
We use the notation $T_\text{bcast}(w,\hat p,\nu,\mu)$ and $T_\text{a2a}(w,\hat p,\nu,\mu)$ to denote the costs of broadcast and all-to-all, where $w$ is the size of the data (per processor) in matrix elements, $\hat p$ is the number of processes participating in the collective, $\nu$ is the number of simultaneous collectives, and $\mu$ is the number of processes per node.
Parameters $\nu$ and $\mu$ capture resource contention: the number of simultaneous collectives affects contention for network bandwidth, and the number of processes per node affects contention for the network interface card on each node.

In general, these cost functions can be approximated via microbenchmarks for a given machine and MPI implementation, though they can vary over different node allocations as well.
If we ignore resource contention, with $\nu=1$ and $\mu=1$, then the costs are typically modeled~\cite{ChanHPG07} as
$$T_\text{bcast}(w,\hat p,1,1) = \alpha \cdot \log{\hat p} + \beta \cdot w \frac{\hat p-1}{\hat p}$$
and 
$$T_\text{a2a}(w,\hat p,1,1) = \alpha \cdot (\hat p-1) + \beta \cdot w \frac{\hat p-1}{\hat p}.$$
The all-to-all cost assumes a point-to-point algorithm, minimizing bandwidth cost at the expense of higher latency cost; see \cite[Section 2.2]{spaa13} for more details on the tradeoffs within all-to-all algorithms.

The time spent in communication is then given by 
\begin{align*}
&T_\text{a2a}\left(\frac{\dnnz(\mB)}{p},c,\frac pc,\mu\right) + \\
\frac{\dimN}{bc} \cdot &T_\text{bcast}\left(\frac b\dimN \cdot \frac{\dnnz(\mA)}{\sqrt{p/c}},\sqrt{p/c},\sqrt{pc},\mu\right)  + \\
\frac{\dimN}{bc} \cdot &T_\text{bcast}\left(\frac b\dimN \cdot \frac{\dnnz(\mB)}{\sqrt{p/c}},\sqrt{p/c},\sqrt{pc},\mu\right)  + \\
&T_\text{a2a}\left(\frac{\flops(\mA, \mB)}{p},c,\frac pc,\mu\right).
\end{align*}
The amount of data communicated in the first three terms is the average over all processes and is accurate only if the nonzeros of the input matrices are evenly distributed across all blocks.
The amount of data communicated in the last term is an upper bound on the average; the number of output matrix entries communicated by each process is likely less than the number of flops performed by that process (due to the reduction of locally repeated indices 
prior to communication). A lower bound for the last term is given by replacing $\flops(\mA, \mB)$ with $\dnnz(\mC)$.

If we ignore resource contention, the communication cost is
\begin{equation*}
\alpha \cdot O\left( \frac{\dimN}{bc} \log(p/c) + c \right) + \beta \cdot O\left( \frac{\dnnz(\mA)+\dnnz(\mB)}{\sqrt{pc}} + \frac{\flops(\mA, \mB)}{p}\right), 
\end{equation*}
where we have assumed that $\dnnz(\mB)\leq \flops(\mA, \mB)$.
Note that this expression matches the costs for \erdosrenyi\ matrices,
up to the choice of all-to-all algorithm~\cite{spaa13}, 
where $\dnnz(\mA)\approx\dnnz(\mB)\approx dn$ and $\flops(\mA, \mB)\approx d^2n$.

We make several observations based on this analysis of the communication.
First, increasing $c$ (the number of layers) results in less time spent in broadcast collectives and more time spent in all-to-all collectives (note that if $c=1$ then no communication occurs in all-to-all).
Second, increasing $b$ (the blocking parameter) results in fewer collective calls but the same amount of data communicated; thus, $b$ navigates a tradeoff between latency cost and local memory requirements (as well as greater possibility to overlap local computation and communication).
Third, for a fixed number of cores, lower $\mu$ (higher value of $t$) will decrease network interface card contention and therefore decrease communication time overall.


\section{Experimental Results}
\label{sec:expparallel}

We evaluate our algorithms on two supercomputers: Cray XC30 at NERSC (Edison)~\cite{Edison_website}, and Cray XK6 at ORNL (Titan)~\cite{Titan_website}. 
Architectural details of these computers are listed in Table~\ref{tab:machines}. In our experiments, we ran only on the CPUs and did not utilize Titan's GPU accelerators.

In both supercomputers, we used Cray's MPI implementation, which
is based on MPICH2. Both chip architectures achieve memory parallelism via hardware prefetching. 
On Titan, we compiled our code using GCC \Cpp\ compiler version 4.6.2 with \texttt{\mbox{\small -O2 -fopenmp}} flags.  
On Edison, we compiled our code using the Intel \Cpp\ compiler (version 14.0.2) with the options \texttt{\mbox{\small -O2 -no-ipo -openmp}}.
In order to ensure better memory affinity to NUMA nodes of Edison and Titan, we  used \texttt{-cc depth} or \texttt{-cc numa\_node} options when submitting jobs. 
For example, to run the 3D algorithm on a $8{\times}8{\times}4$ process grid with 6 threads, we use the following options on Edison: \texttt{aprun -n 256 -d 6 -N 4 -S 2 -cc depth}. 
In our experiments, we always allocate cores needed for a particular configuration of 3D algorithms, i.e., to run the 3D algorithm on $\sqrt{p/c} \times \sqrt{p/c} \times c$ process grid with $t$ threads per process, we allocate $pt$ cores and run $p$  MPI processes on the allocated cores.

{
\setlength{\tabcolsep}{5pt}
\begin{table}[!tb]{
\centering
\begin{tabular}{rcc}
					&  {\bf Cray XK7 (Titan) } & {\bf Cray XC30 (Edison) }    \\
\hline
{\bf Core}  & {\bf AMD Interlagos}  & {\bf Intel Ivy Bridge}		\\
\hline
Clock (GHz)			& 2.2				& 2.4					\\
Private Cache (KB)		& 16+2048 		& 32+256				\\
DP GFlop/s/core		& 8.8				&19.2		\\
\hline
{\bf Socket Arch.} 	        & {\bf Opteron 6172} & {\bf Xeon E5-2695 v2}	\\
\hline
Cores per socket			& 16				& 12					\\
Threads per socket		& 16				& 24$^1$				\\
L3 cache per socket		& 2$\times$8~MB	&	30~MB			\\
\hline
{\bf Node Arch.}	 	 & 	Hypertransport 		& QPI (8 GT/s)\\
\hline
Sockets/node			&  1		&	2					\\
STREAM BW$^2$		&  31~GB/s 	&	104~GB/s		\\
Memory per node		&  32~GB	&	64~GB			\\
\hline
Interconnect		& Gemini (3D Torus) & Aries (Dragonfly)	\\
\hline
\end{tabular}
\caption{Overview of Evaluated Platforms.  $^1$Only 12 threads were used.  $^2$Memory bandwidth is measured using the STREAM copy benchmark per node.}
\label{tab:machines}
}
\end{table}
}

\begin{figure}[!t]
\begin{center}

\renewcommand{\arraystretch}{1.2}
\addtolength{\tabcolsep}{-2.6pt}
\begin{tabular}{|lccc|}
	\hline 
\textbf{Name}		& \multirow{3}{*}{Spy Plot} 				& Dimensions		& nnz/row 		\\ 
Description		&							& Nonzeros		& symmetric	\\
	\hline
			& \multirow{3}{*}{\includegraphics[scale=0.11]{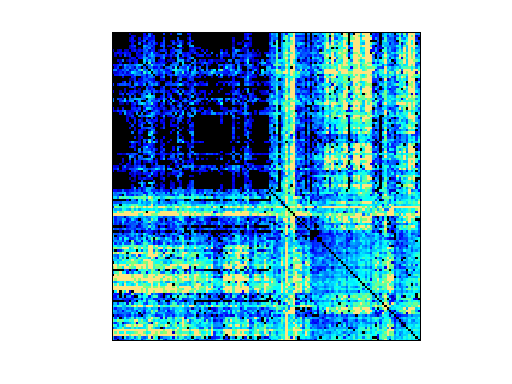}}& 			&			\\ 
\textbf{mouse\_gene}	&  							&  45K$\times$45K	& 	642	\\
Gene network 	&							&  28.9M		& 	\checkmark	\\	
	\hline
				& \multirow{3}{*}{\includegraphics[scale=0.12]{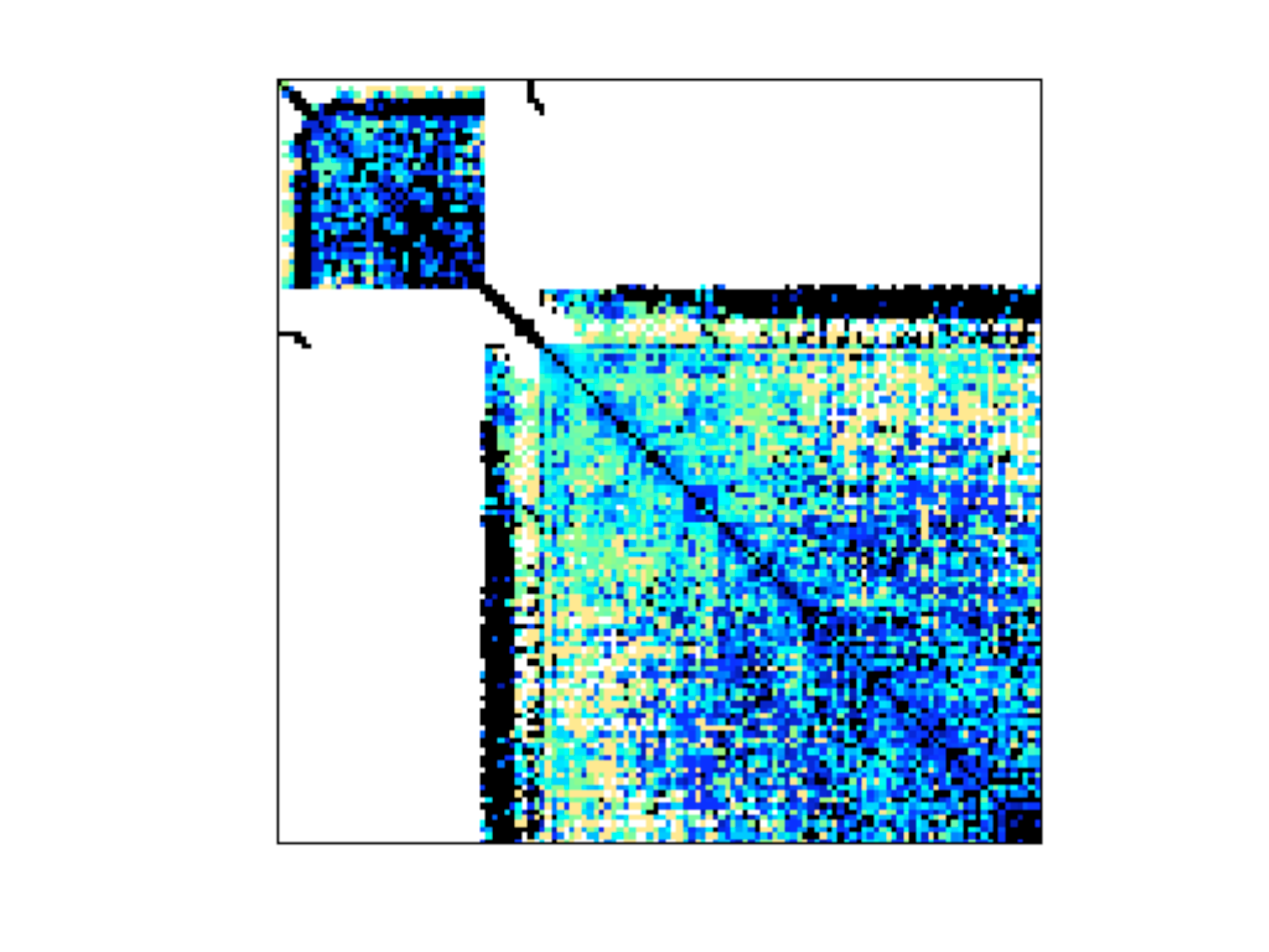}}	& 			&			\\
\textbf{ldoor}	& 								& 952K$\times$952K	& 	48.8 		\\
structural problem	&							& 46.5M		& 	\checkmark	\\
	\hline
			& \multirow{3}{*}{\includegraphics[scale=0.11]{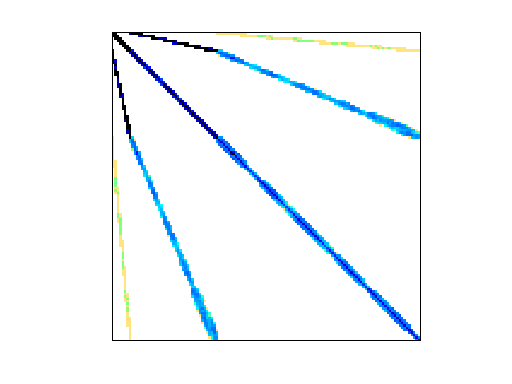}}& 			&			\\
\textbf{dielFilterV3real}	& 							& 1.1M$\times$1.1M	& 	81.2 		\\
electromagnetics problem		&							& 89.3M		& 	\checkmark	\\
	\hline	
			&  \multirow{3}{*}{\includegraphics[scale=0.11]{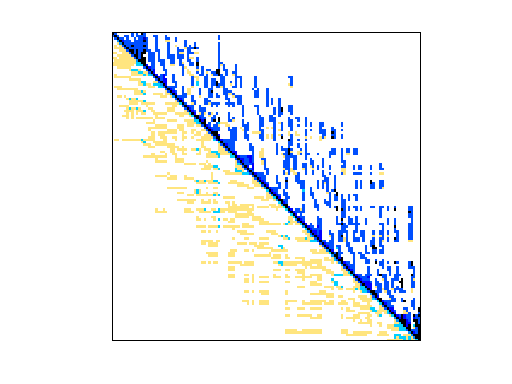}}	& 			& 			\\
\textbf{cage15}		&							& 5.15M$\times$5.15M	& 	19.3	 	\\
DNA electrophoresis 		&							& 99.2M		& 	 	\\
	\hline
			& \multirow{3}{*}{\includegraphics[scale=0.11]{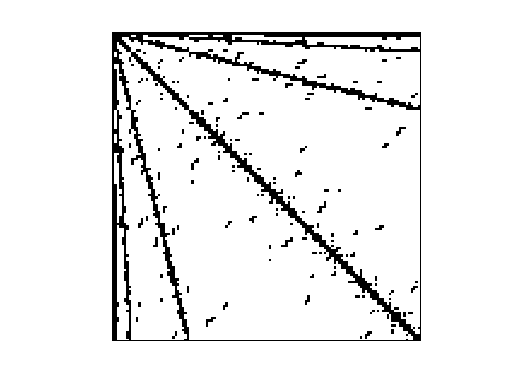}}	& 			&			\\
\textbf{delaunay\_n24}		&	 						& 16.77M$\times$16.77M		& 	6 	\\
Delaunay triangulation		&							& 100.6M			& 	\checkmark		\\
	\hline
			& \multirow{3}{*}{\includegraphics[scale=0.11]{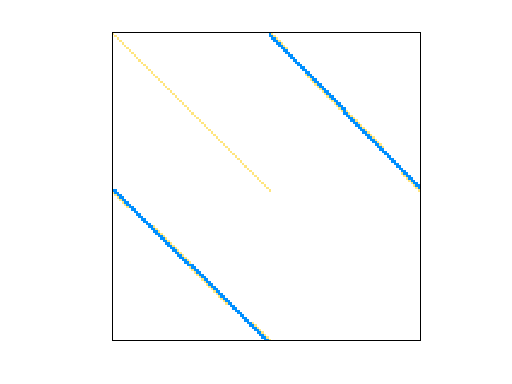}}	& 			& 			\\
\textbf{nlpkkt160}	& 							& 8.34M$\times$8.34M	& 	27.5		 \\ 
indefinite KKT matrix			&							& 229.5M		& 	 \checkmark	\\
	\hline

			& \multirow{3}{*}{\includegraphics[scale=0.11]{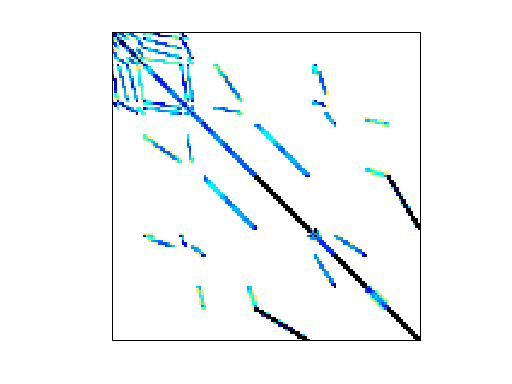}}	&			&			\\
\textbf{HV15R}	& 							& 2.01M$\times$2.01	& 	141.5		\\
3D engine fan 	&								& 283M		& 	\\
	\hline
				& \multirow{3}{*}{\includegraphics[scale=0.11]{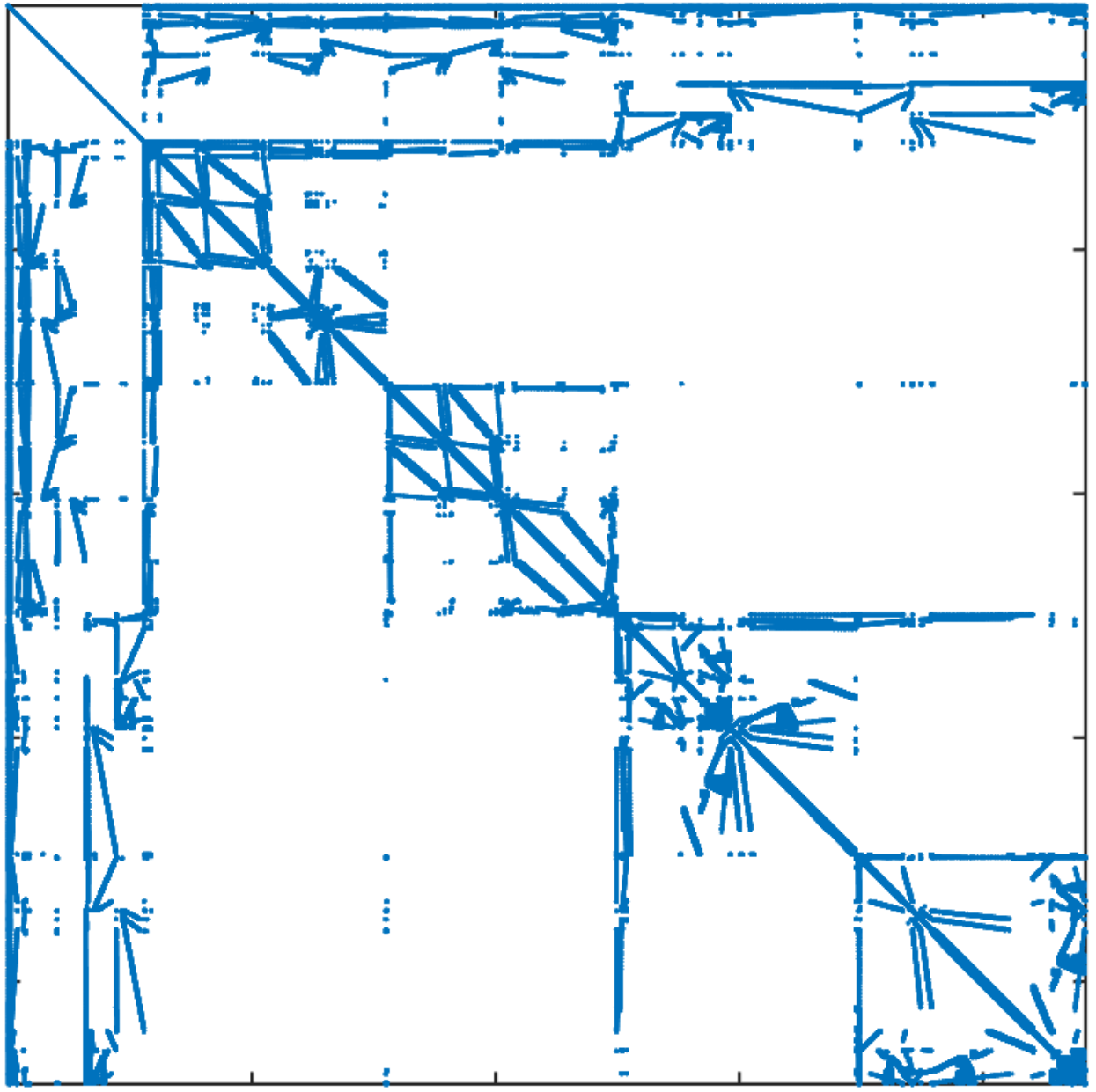}}	&			&			\\
\textbf{NaluR3}	& 							& 17.6M$\times$17.6M	& 26.9		\\
Low Mach fluid flow 	&								& 474M		& 	\checkmark	\\
	\hline
					& \multirow{3}{*}{\includegraphics[scale=0.11]{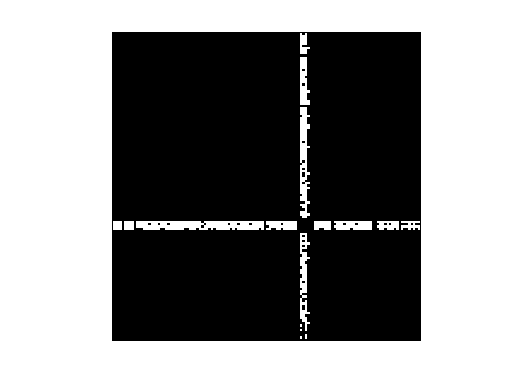}}	&			&			\\
\textbf{it-2004}	& 							& 41.29M$\times$41.29M	& 27.8		\\
web crawl of .it domain 	&								& 1,150M		& 		\\
	\hline
\end{tabular}
\end{center}
\caption{\small \bf Structural information on the sparse matrices used in our
  experiments.  All matrices 
  are from the University of Florida sparse
  matrix collection~\cite{davis2011university}, except NaluR3, which is a matrix from low Mach number, turbulent
reacting flow problem~\cite{lin2014towards}. 
For the Florida matrices, we consider the explicit zero entries to be nonzeros and update the nnz of the matrices accordingly.
}
\label{fig:testsuite}
\vsf
\end{figure}

Several software libraries support SpGEMM. For GPUs, CUSP and CUSparse implement SpGEMM. 
For shared-memory nodes, MKL implements SpGEMM.
Trilinos package implements distributed memory SpGEMM~\cite{Her2005}, which uses a 1D decomposition for its sparse matrices.
In this paper, we compared the performance of 2D and 3D algorithms with SpGEMM in Cray-Trilinos package (version 11.6.1.0)  available in NERSC computers,
which features significant performance improvements over earlier versions.
Sparse SUMMA is the 2D algorithm that had been published before~\cite{icpp08} without in-node multithreading, and \SpGEMMThreeD{} is the 3D algorithm first presented here.
Sometimes we will drop the long names and just use 2D and 3D for abbreviation.


In our experiments, we used both synthetically generated matrices, as well as real matrices from several different sources. 
In Section \ref{sec:square_exp}, we benchmark square matrix multiplication.
We use \rmat~\cite{rmat}, the Recursive MATrix generator to generate three different classes of synthetic matrices: (a) G500 matrices  representing graphs with skewed degree distributions from Graph 500 benchmark~\cite{graph500}, (b) SSCA matrices from HPCS Scalable Synthetic Compact Applications graph analysis (SSCA\#2) benchmark~\cite{SSCA}, and (c)  ER matrices representing \erdosrenyi~random graphs.
We use the following \rmat\  seed parameters to generate these matrices: (a) $a=.57$, $b=c=.19$, and $d=.05$ for G500, (b) $a=.6$, and $b=c=d=.4/3$ for SSCA, and (c) $a=b=c=d=.25$ for ER.
A scale $n$ synthetic matrix is $2^n$-by-$2^n$.
On average, G500 and ER matrices have $16$ nonzeros, and SSCA matrices have $8$ nonzeros per row and column.
We applied a random symmetric permutation to the input matrices
to balance the memory and the computational load.
In other words, instead of storing and computing $\mC = \mA \mB$, 
we compute $\mathbf{P} \mC \mathbf{P}\transpose = (\mathbf{P} \mA \mathbf{P}\transpose )(\mathbf{P} \mB \mathbf{P}\transpose)$.
All of our experiments are performed on double-precision floating-point inputs, and matrix indices are stored as 64-bit integers.

In Section~\ref{sec:exp_restriction}, we benchmark the matrix multiplication corresponding to the restriction operation that is used in AMG.
Since AMG on graphs coming from physical problems is an important case, 
we include several matrices from the Florida Sparse Matrix collection~\cite{davis2011university} in our experimental analysis.
In addition, since AMG restriction is computationally isomorphic to the graph contraction operation performed by multilevel graph partitioners~\cite{brucegraph95}, 
we include a few matrices representing real-world graphs.  

The characteristics of the real test matrices are shown in Table~\ref{fig:testsuite}.
Statistics about squaring real matrices and multiplying each matrix with its restriction operator $\mR$ is given in Table~\ref{tab:stats}.

\begin{table}[!t]
   \centering
   \topcaption{Statistics about squaring real matrices and multiplying each matrix with its restriction operator $\mR$.} 
            \scalebox{0.9}{
   \begin{tabular}{@{} l r r r r r @{}} 
      \toprule
     Matrix ($\mA$)    & $nnz(\mA)$ & $nnz(\mA^2)$ & $nnz(\mR)$ & $nnz(\mR\transpose\mA)$  &$nnz(\mR\transpose\mA\mR)$ \\
      \midrule
      \textbf{mouse\_gene}      & 28,967,291  & 482,594,045 &  45,101 & 2,904,560 & 402,200\\
      \textbf{ldoor}      & 46,522,475  & 145,422,935 &  952,203 & 2,308,794 & 118,093\\
       \textbf{dielFilterV3real}      & 89,306,020  & 688,649,400 &  1,102,824 & 4,316,781 & 100,126\\
      \textbf{cage15}      & 99,199,551  & 929,023,247 &  5,154,859 & 46,979,396 & 17,362,065\\
      \textbf{delaunay\_n24}      & 100,663,202  & 347,322,258 &  16,777,216 & 41,188,184 & 15,813,983 \\
      \textbf{nlpkkt160}      & 229,518,112  & 1,241,294,184 &  8,345,600 & 45,153,930 & 3,645,423\\
      \textbf{HV15R}      & 283,073,458  & 1,768,066,720 &  2,017,169 & 10,257,519 & 1,400,666\\
       \textbf{NaluR3}      & 473,712,505  & 2,187,662,967 &  17,598,889 & 77,245,697 & 7,415,297\\
      \textbf{it-2004}      & 1,150,725,436  & 14,045,664,641 &  41,291,594 & 89,870,859 & 26,847,490\\
      \bottomrule
   \end{tabular}}
   \label{tab:stats}
   \vsf
\end{table}

\subsection{Intra-node Performance} 
Our 3D algorithm exploits intra-node parallelism in two computationally intensive functions: (a) local \proc{HeapSpGEMM} performed by each MPI process at every SUMMA stage, and (b) multiway merge performed at the end of all SUMMA stages.
As mentioned before, \proc{HeapSpGEMM} returns a set of intermediate triples that are 
kept in memory and merged at the end of all SUMMA stages.
In this section, we only show the intra-node scalability of these two functions and compare them against an MKL and a GNU routine.

\subsubsection{Multithreaded \proc{HeapSpGEMM} Performance}
\label{sec:perfmtspgemm}
\begin{figure}[!t]
   \centering
   \subfloat[][Scale 16 G500 $\times$ G500]{\includegraphics[scale=.44]{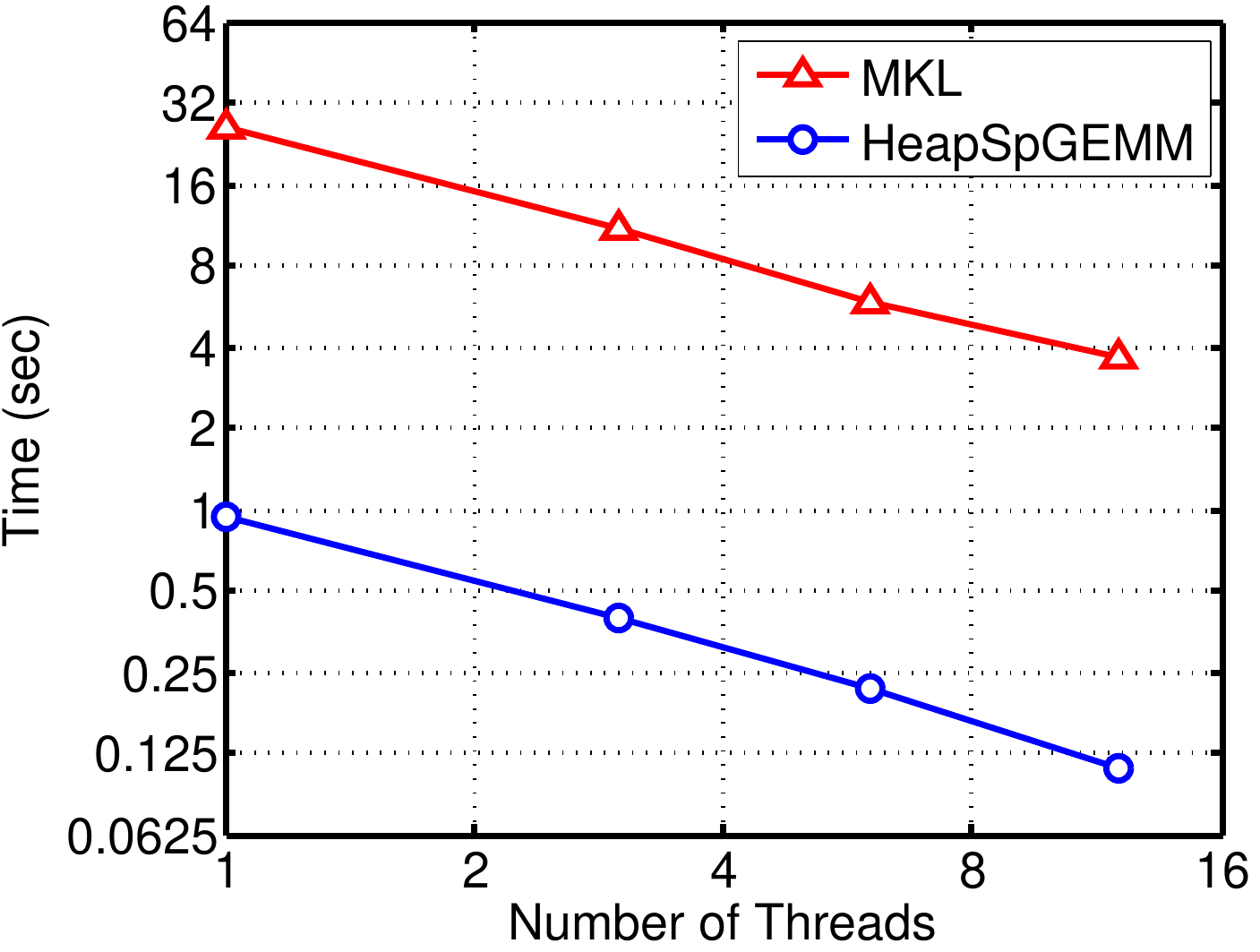}}
   ~
   \subfloat[][cage12  $\times$ cage12]{\includegraphics[scale=.44]{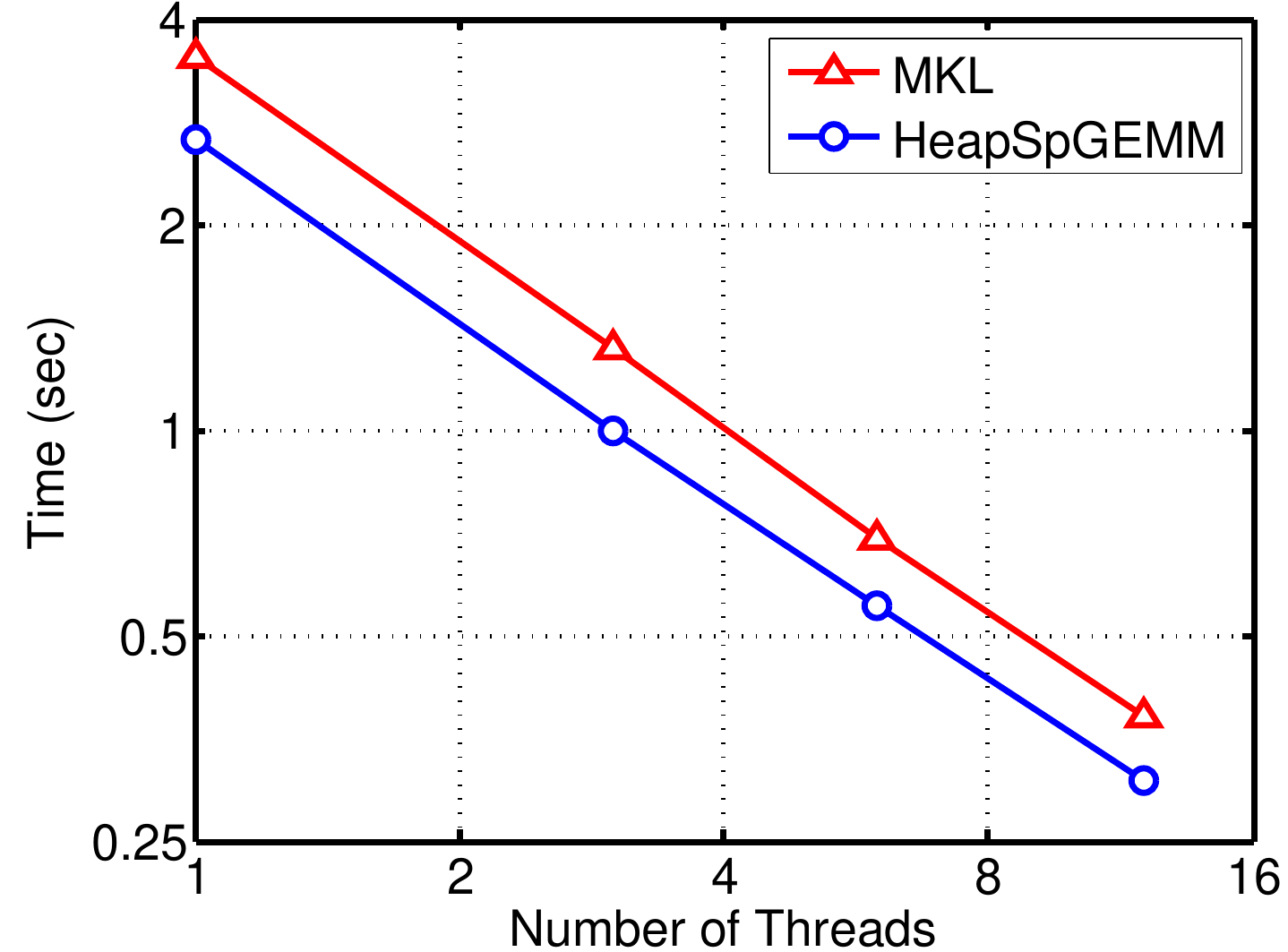}} 
   \caption{Thread scaling of our \proc{HeapSpGEMM} and the MKL routine \texttt{mkl\_csrmultcsr} on 1,3,6,12 threads (with column indices sorted in the increasing order for each row) when squaring a matrix on a single socket of Edison.}
   \label{fig:local-SpGEMM-scaling}
    \vsf
\end{figure}

We study intra-node scalability of local SpGEMM by running a single MPI process on a socket of a node and varying the number of threads from one to the maximum number of threads available in a socket. We compare the performance of \proc{HeapSpGEMM} with MKL routine \texttt{mkl\_csrmultcsr}. 
To expedite the multiway merge that is called on the output of \proc{HeapSpGEMM}, we always keep column indices sorted in increasing order within each row.
Hence, we ask \texttt{mkl\_csrmultcsr} to return sorted output.
We show the performance of  \proc{HeapSpGEMM} and MKL in Figure~\ref{fig:local-SpGEMM-scaling} where these functions are used to multiply (a) two randomly generated scale 16 G500 matrices, and (b) Cage12 matrix by itself. 
On 12 threads of Edison, \proc{HeapSpGEMM} achieves $8.5\times$ speedup for scale 16 G500 and $8.7\times$ speedup for Cage12, whereas MKL achieves $7.1\times$ speedup for scale 16 G500 and $9.2\times$ speed for Cage12.
Hence, \proc{HeapSpGEMM} scales as well as MKL.
However, for these matrices, \proc{HeapSpGEMM} runs faster than MKL on any concurrency with up to 33-fold performance improvement for G500 matrices.

\begin{figure}[!t]
   \centering
   \subfloat[][4-way merge]{\includegraphics[scale=.43]{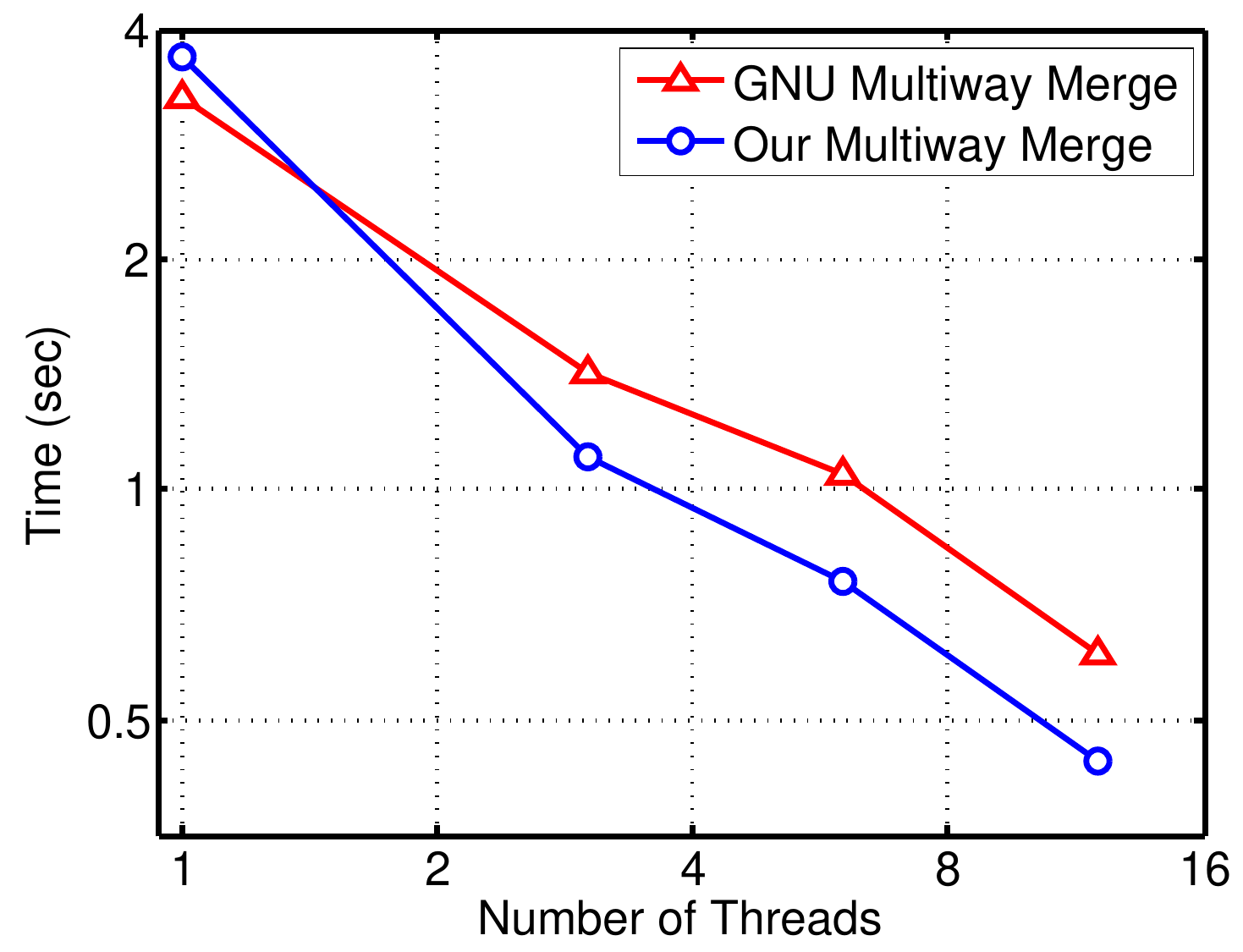}}
   ~
   \subfloat[][16-way merge]{\includegraphics[scale=.43]{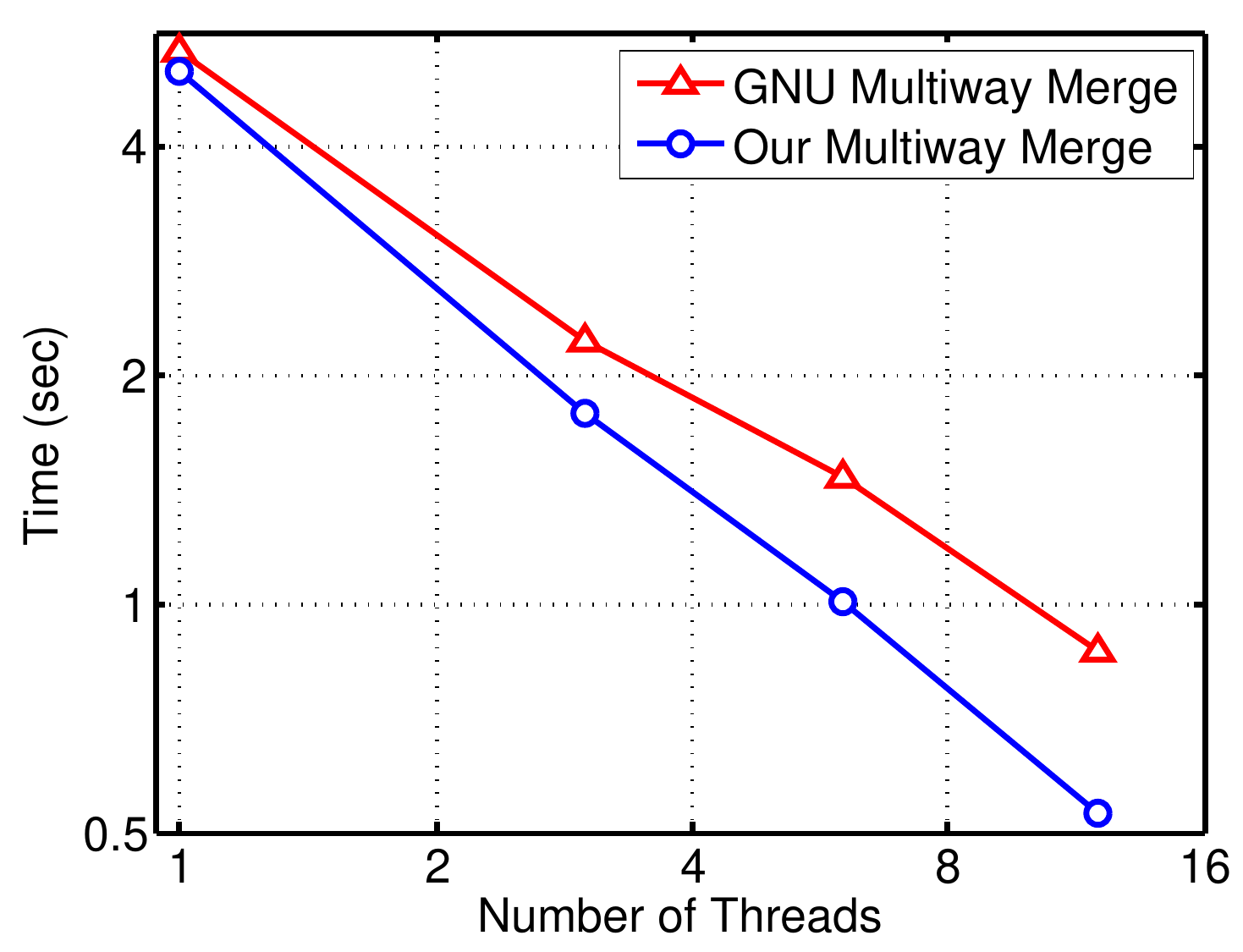}} 
   \caption{Thread scaling of our multiway merge and GNU multiway merge routine \texttt{\_\_gnu\_parallel::multiway\_merge} augmented with a procedure that reduces repeated indices: (a) in squaring scale 21 G500 matrix on 4$\times$4 process grid and (b) in squaring scale 26 G500 matrix on 16$\times$16 process grid on Edison.
   }
   \label{fig:local-merge-scaling}
    \vsf
\end{figure}

\subsubsection{Multithreaded Multiway Merge Performance}
On a $\sqrt{p/c} \times \sqrt{p/c} \times c$ process grid, each MPI process performs two multiway-merge operations. The first one merges $\sqrt{p/c}$  lists of triples computed in  $\sqrt{p/c}$ stages of SUMMA and the second one merges $c$ lists of triples after splitting the previously merged list across layers.
Since both merges are performed by the same function, we experimented the intra-node performance of  multiway merge on a single layer ($c{=}1$).
For this experiment, we allocate $12p$ cores on Edison and run SUMMA on a $\sqrt{p} \times \sqrt{p}$ process grid. Each MPI process is run on a socket using up to 12 available threads.  
Figure~\ref{fig:local-merge-scaling} shows the merge time needed by MPI rank 0 for a 4-way merge and a 16-way merge when multiplying two G500 matrices on a $4 \times 4$ and a $16 \times 16$ grid, respectively.
We compare the performance of our multiway merge with a GNU multiway merge routine \texttt{\_\_gnu\_parallel::multiway\_merge}.
However, the latter merge routine simply merges lists of triples keeping them sorted by column and row indices, but does not reduce triples with the same (row, column) pair. 
Hence, we reduce the repeated indices returned by \texttt{\_\_gnu\_parallel::multiway\_merge} by a multithreaded reduction function and report the total runtime.
From Figure~\ref{fig:local-merge-scaling} we observe that our routine performs both 4-way and 16-way merges faster than augmented GNU multiway merge for G500 matrices.
On 12 threads of Edison, our multiway merge attains $8.3\times$ speedup for 4-way merge and $9.5\times$ speedup for 16-way merge.
By contrast,  the augmented GNU merge attains $5.4\times$ and $6.2\times$ speedups for 4-way and 16-way merges, respectively. 
We observe similar performances for other matrices as well.


\subsection{Square Sparse Matrix Multiplication}
\label{sec:square_exp}
In the first set of experiments, we square real matrices from Table~\ref{fig:testsuite} and multiply two structurally similar randomly generated matrices. 
This square multiplication is representative of the expansion operation used in the Markov clustering algorithm~\cite{vandongen00}. 
We explore an extensive set of parameters of Sparse SUMMA (2D) and \SpGEMMThreeD{} (which is the main focus of this work), identify optimum parameters on difference levels of concurrency, empirically explain where \SpGEMMThreeD{} gains performance, and then show the scalability of \SpGEMMThreeD{} for a comprehensive set of matrices.

\begin{figure}[!t]
   \centering
   \includegraphics[scale=.55]{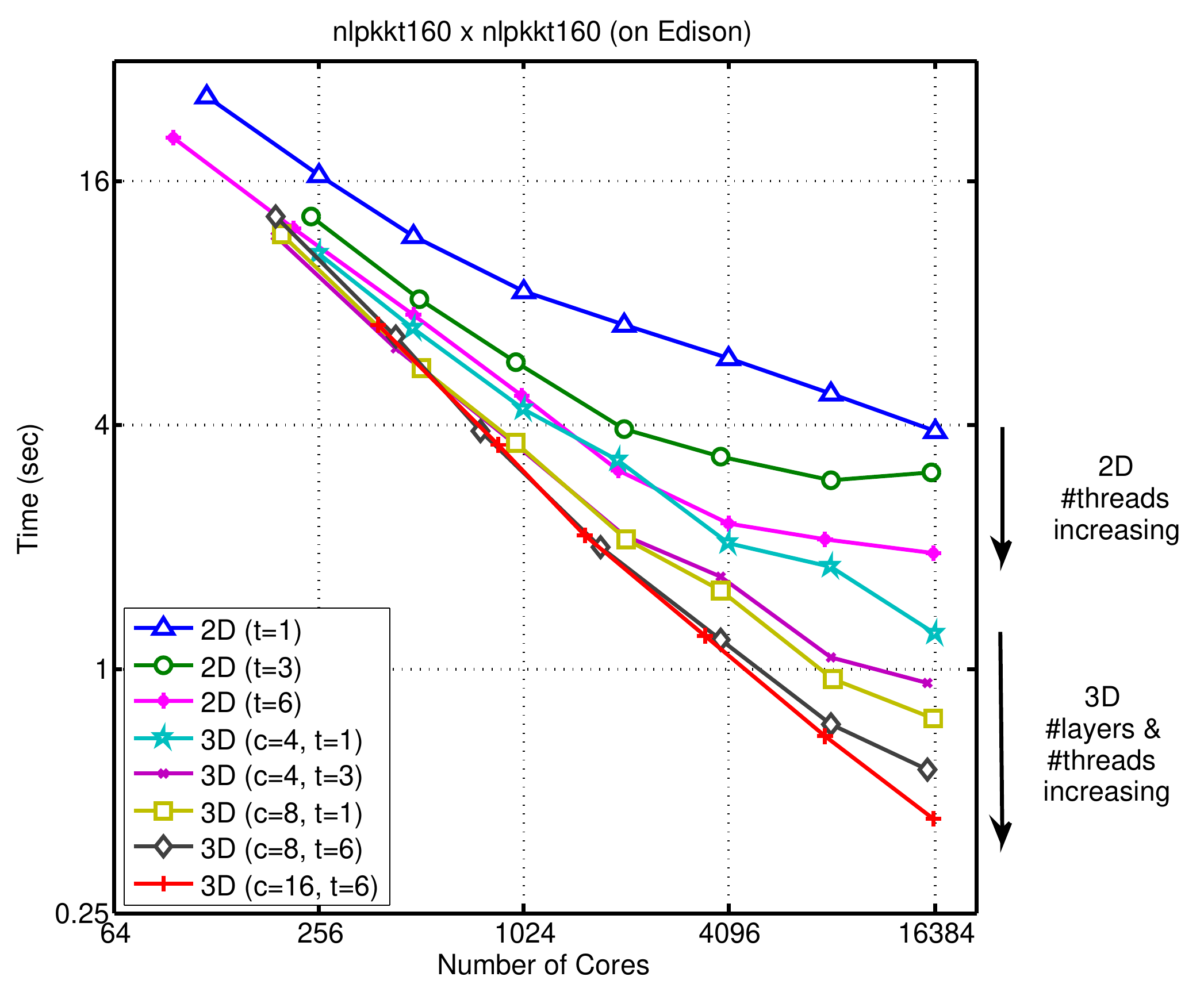} 
   \caption{Strong scaling of different variants of 2D (Sparse SUMMA) and 3D (\SpGEMMThreeD{}) algorithms when squaring of \texttt{nlpkkt160} matrix on Edison. Performance benefits of the 3D algorithm and multithreading can be realized on higher concurrency.
   2D non-threaded algorithm attains about $50\times$ speedup when we go from 1 core to 256 cores, and  2D algorithm with 6 threads attains about $25\times$ speedup when we go from 6 cores to 216 cores (not shown in the figure). }
   \label{fig:nlpkkt160-strong-scaling}
    \vsf
\end{figure}

\begin{figure}[!t]
   \centering
   \includegraphics[scale=.54]{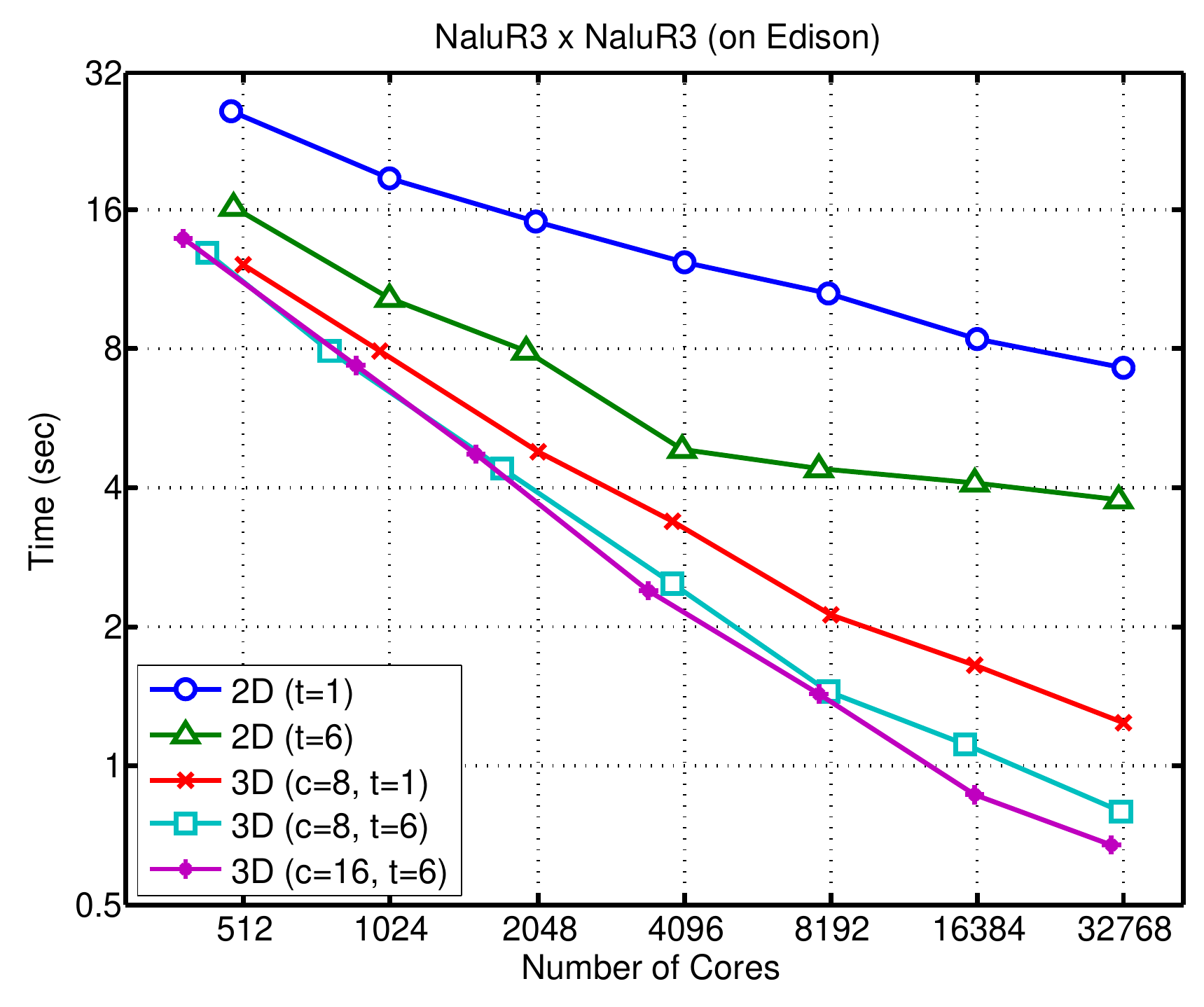} 
   \caption{Strong scaling of different variants of 2D and 3D algorithms when squaring of \texttt{NaluR3} matrix on Edison. 3D algorithms are an order of magnitude faster than 2D algorithms on higher concurrency.}
   \label{fig:NaluR3_AA}
    \vsf
\end{figure}

\begin{figure}[!t]
   \centering
   \includegraphics[scale=.54]{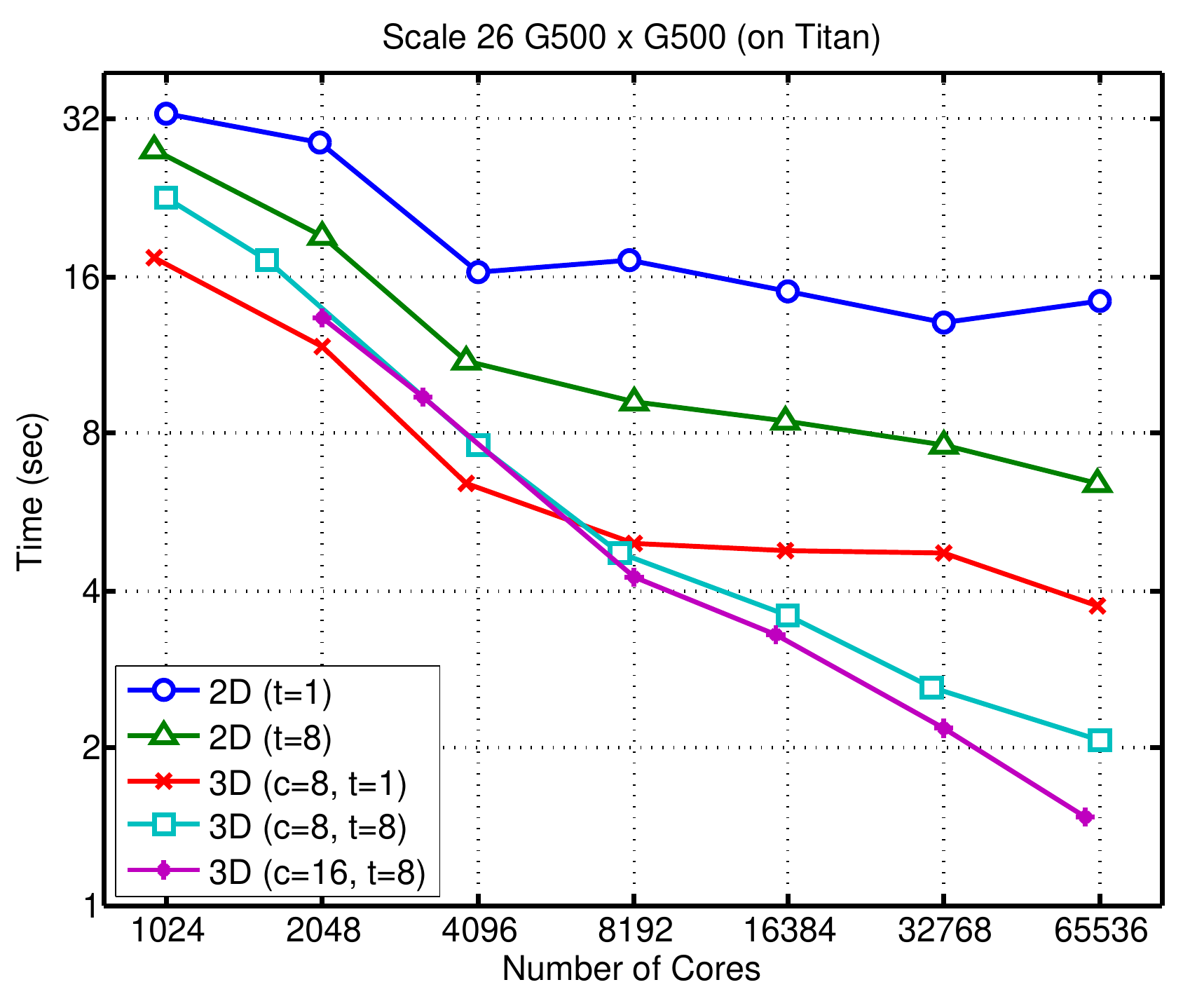} 
   \caption{Strong scaling of different variants of 2D and 3D algorithms on Titan when multiplying two scale 26 G500 matrices. 3D algorithm and multithreading improve performance of SpGEMM on higher concurrency}
   \label{fig:G500-26-strong-scaling}
   \vsf
\end{figure}

\subsubsection{Performance of Different Variants of 2D  and 3D Algorithms}
At first, we investigate the impact of multithreading and 3D algorithm on the performance of SpGEMM.
For this purpose, we fix the number of cores $p$ and multiply two sparse matrices with different combinations of thread counts $t$ and number of layers $c$.
Figure~\ref{fig:nlpkkt160-strong-scaling} shows strong scaling of squaring of \texttt{nlpkkt160} matrix on Edison.
On lower concurrency ($<256$ cores), multithreading improves the performance of 2D algorithm, e.g., about $1.5\times$ performance improvement with 6 threads on 256 cores in Figure~\ref{fig:nlpkkt160-strong-scaling}.
However, there is little or no benefit in using a 3D algorithm over a multithreaded 2D algorithm on lower concurrency because the processor grid in a layer becomes too small for 3D algorithms.

For better resolution on higher concurrency, we have not shown the runtime of 2D algorithms before 64 cores in Figure~\ref{fig:nlpkkt160-strong-scaling}. 
For the completeness of our discussion, we briefly discuss the performance of our algorithm on lower concurrency and compare them against MKL and Matlab.
Matlab uses an efficient CSC implementation of Gustavson's algorithm.
2D non-threaded algorithm takes about  800 seconds on a single core and attains about $50\times$ speedup when we go from 1 core to 256 cores, and 2D algorithm with 6 threads attains about $25\times$ speedup when we go from 6 cores to 216 cores.
By contrast, on a single core, MKL and Matlab take about 500 and 830 seconds, respectively to square randomly permuted \texttt{nlpkkt160} matrix (in Matlab, we keep explicit zero entries\footnote{The default Matlab behavior is to remove entries with zero values when constructing a matrix using its sparse(i,j,v)} to obtain the same number of nonzeros shown in Table~\ref{tab:stats}).
Therefore, the serial performance of Sparse SUMMA (2D) is comparable to that of MKL and Matlab.
The best single node performance is obtained by multithreaded SpGEMM.
Using 24 threads on 24 cores of a single node of Edison, MKL and HeapSPGEMM take about 32 and 30 seconds, respectively.
We note that the above performance numbers depend significantly on nonzero structures of the input matrices.
Here, we select  \texttt{nlpkkt160} matrix for discussion because the number of nonzero in the square of \texttt{nlpkkt160}  is about 1.2 billion (c.f. Table~\ref{tab:stats}), requiring about 28GB of memory to store the result, which is close to the available single node memory of Edison.

The performance benefits of the 3D algorithm and multithreading become more dominant on higher concurrency.
In Figure~\ref{fig:nlpkkt160-strong-scaling}, when we increase $p$ from 256 to 16,384 ($64\times$ increase), non-threaded 2D and 3D ($c{=}16, t{=}6$) algorithms run $4\times$ and $22\times$ faster, respectively.
Consequently, on 16,384 cores, \SpGEMMThreeD{} with $c{=}16, t{=}6$ multiplies \texttt{nlpkkt160} matrix $8\times$ faster than non-threaded 2D algorithm.
We observe similar trend for other real and randomly generated matrices as well.
For example, \SpGEMMThreeD{} with $c{=}16, t{=}6$  runs  $10\times$ faster than the Sparse SUMMA (2D) algorithm when squaring of NaluR3 on  32,764 cores of Edison (Figure~\ref{fig:NaluR3_AA}), and \SpGEMMThreeD{} with $c{=}16, t{=}8$  runs  $9.5\times$ faster than 2D algorithm when multiplying two scale 26 RMAT matrices on 65,536 cores of Titan (Figure~\ref{fig:G500-26-strong-scaling}).

In fact, on higher concurrency, the time \SpGEMMThreeD{} takes to multiply two square matrices decreases gradually with the increase of $c$ and $t$ as indicated on the right side of Figure~\ref{fig:nlpkkt160-strong-scaling}.
This trend is also observed in Figures~\ref{fig:NaluR3_AA} and \ref{fig:G500-26-strong-scaling}.
Therefore, we expect that using more threads and layers will be beneficial to gain performance on even higher concurrency.

\begin{figure}[!t]
   \centering
   \includegraphics[scale=.5]{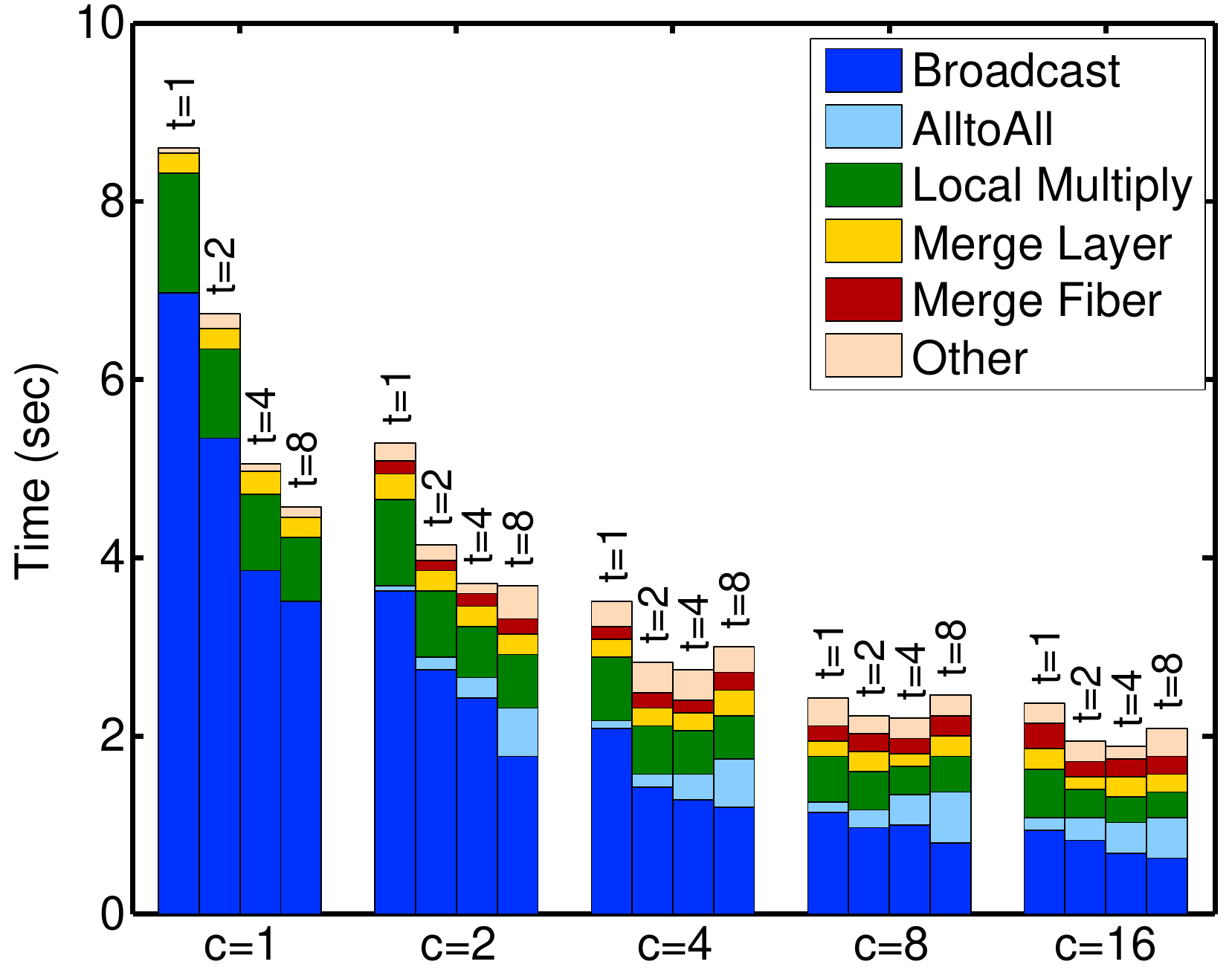} 
   \caption{Breakdown of runtime spent by \SpGEMMThreeD{} for various ($c, t$) configurations on 8,192 cores of Titan when multiplying two scale 26 G500 matrices. The broadcast time (the most dominating term on high concurrency) decreases gradually with the increase of both $c$ and $t$, which is the primary catalyst behind the improved performance of multithreaded 3D algorithms.}
   \label{fig:Timebreak-Scale26-G500-8192}
    \vsf
\end{figure}

\begin{figure}[!t]
   \centering
   \includegraphics[scale=.5]{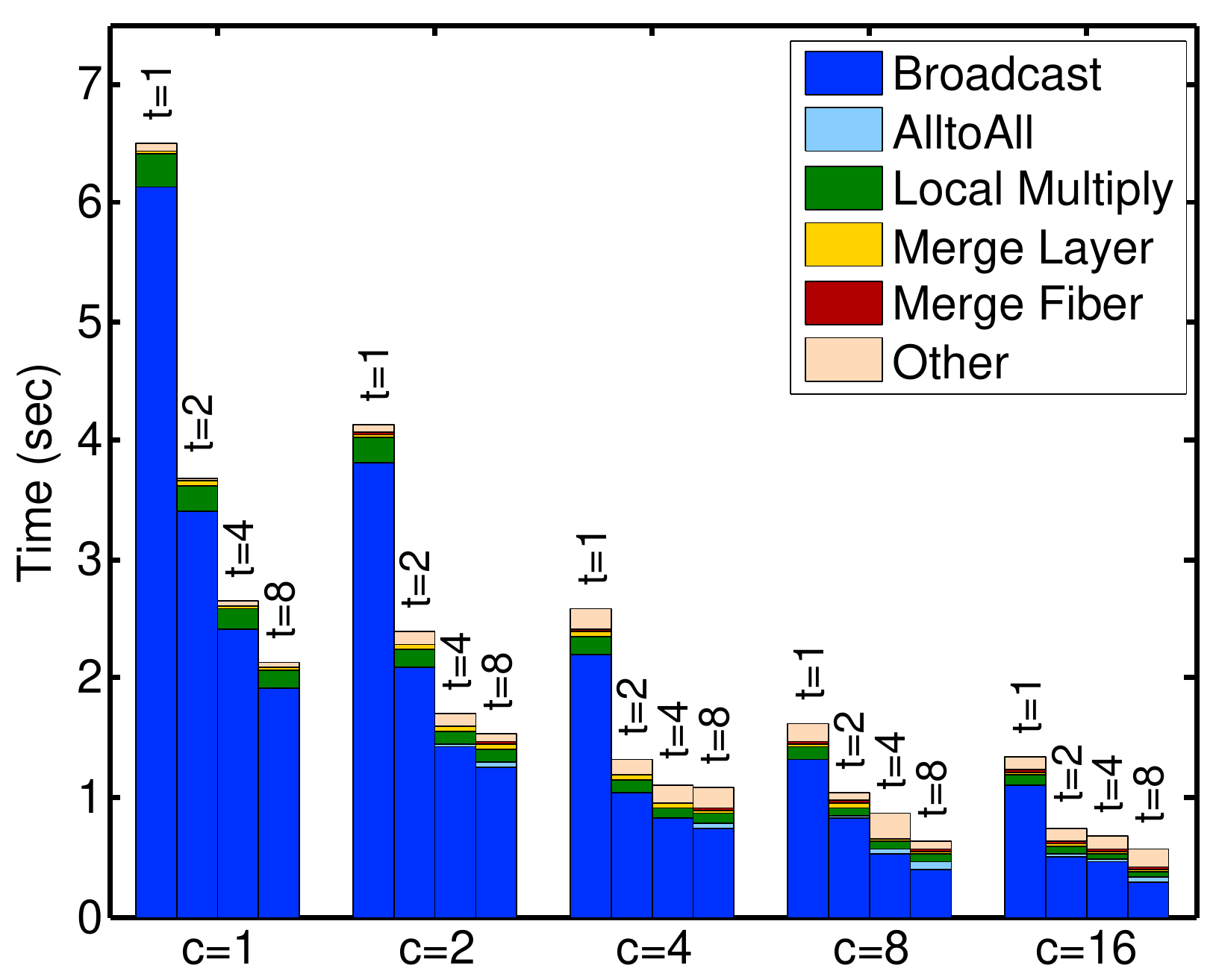} 
   \caption{Breakdown of runtime spent by \SpGEMMThreeD{} for various ($c, t$) configurations on 32,768 cores of Titan when multiplying two scale 26 G500 matrices.}
   \label{fig:Timebreak-Scale26-G500-32768}
    \vsf
\end{figure}

\subsubsection{Breakdown of Runtime}
To understand the performance of \SpGEMMThreeD{}, we break down the time spent in communication and computation when multiplying two G500 graphs of scale 26 and show them in Figure~\ref{fig:Timebreak-Scale26-G500-8192} for 8,192 cores and Figure~\ref{fig:Timebreak-Scale26-G500-32768} for 32,768 cores on Titan.
Here, ``Broadcast" refers to the time needed to broadcast pieces of $\mA$ and $\mB$ within each layer, ``AlltoAll" refers to the communication time needed to communicate pieces of $\mC$ across layers, ``Local Multiply" is the time needed by multithreaded \proc{HeapSpGEMM}, ``Merge Layer" is the time to merge $\sqrt{p/c}$  lists of triples computed in  $\sqrt{p/c}$ stages of SUMMA within a layer, and ``Merge Fiber" is the time to merge $c$ lists of triples after splitting pieces of $\mC$ across processor fibers.
For a fixed number of cores, the broadcast time gradually decreases with the increase of both $c$ and $t$, because as we increase $c$ and/or $t$, the number of MPI processes participating in broadcast within each process layer decreases.
For example, in Figure~\ref{fig:Timebreak-Scale26-G500-32768}, the broadcast time decreases by more than $5\times$ from the leftmost bar to the rightmost bar. 
Since broadcast is the dominating term on higher concurrency, reducing it improves the overall performance of SpGEMM.
However, for a fixed number of cores, the All2All time increases with  $c$ due to the increased processor count on the fiber. 
The All2All time also increases with $t$ because  each MPI process owns a bigger portion of the data, increasing the All2All communication cost per process. 
Therefore, increased All2All time might nullify the advantage of reduced broadcast time when we increase $c$ and $t$, especially on lower concurrency.
For example, using $c{>}4$ does not reduce the total communication time on  8,192 as shown in Figure~\ref{fig:Timebreak-Scale26-G500-8192}.

\begin{figure}[!t]
   \centering
   \subfloat[][\texttt{NaluR3}  $\times$ \texttt{NaluR3}  ($c=1, t=1$)]{\includegraphics[scale=.3]{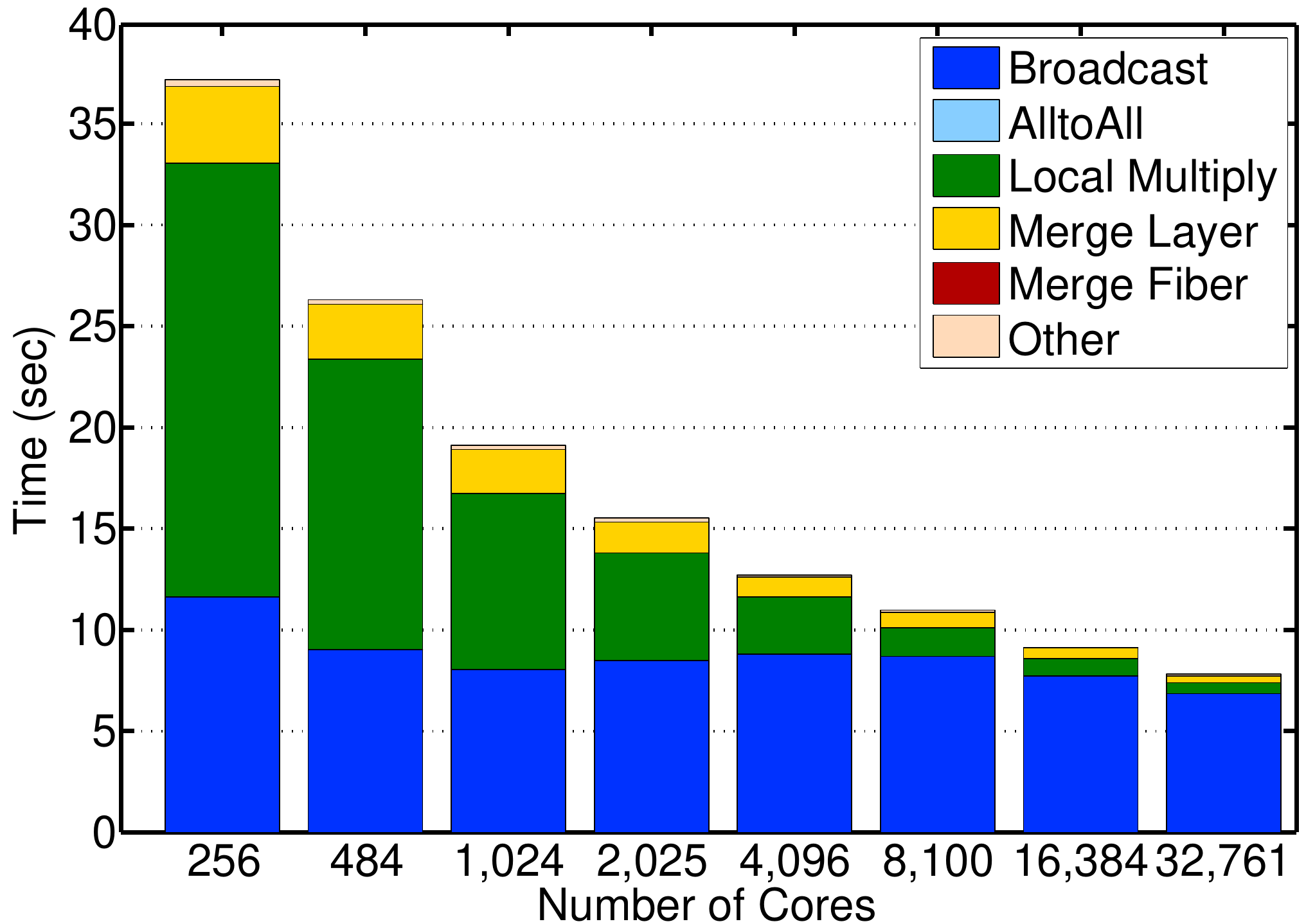}}
   ~~
   \subfloat[][\texttt{NaluR3}  $\times$ \texttt{NaluR3}  ($c=16, t=6$)]{\includegraphics[scale=.3]{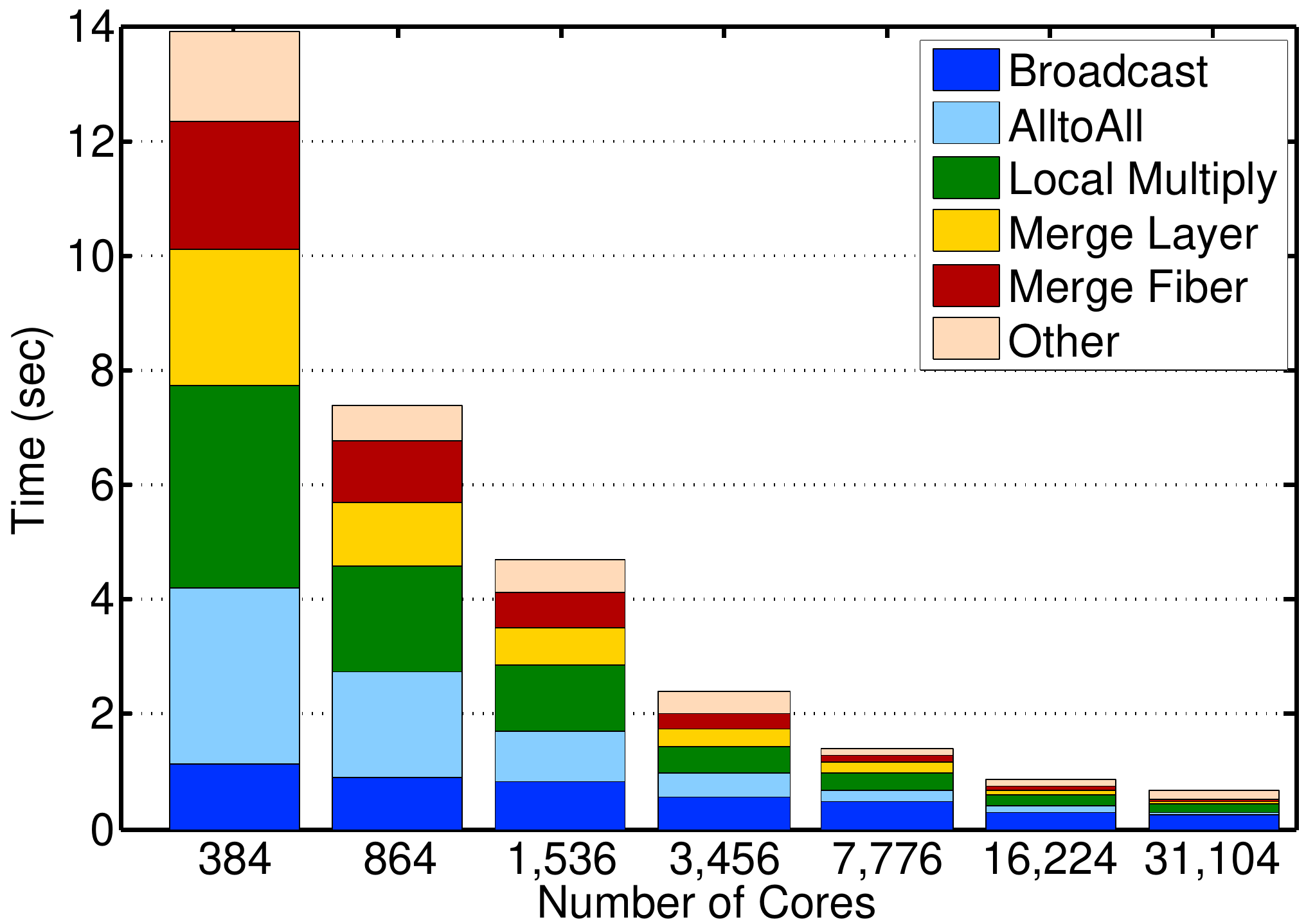}} 
   \caption{Breakdown of runtime spent by (a) Sparse SUMMA (2D) algorithm with $c=1, t=1$ and (b) \SpGEMMThreeD{} algorithm with $c=16, t=6$ to square \texttt{NaluR3} on Edison.}
   \label{fig:scaleBarNaluR3}
    \vsf
\end{figure}

Figure~\ref{fig:Timebreak-Scale26-G500-8192} and Figure~\ref{fig:Timebreak-Scale26-G500-32768} reveal that shorter communication time needed by \SpGEMMThreeD{} makes it faster 
than Sparse SUMMA (2D) on higher concurrency.
Figure~\ref{fig:scaleBarNaluR3}(b) demonstrates that both communication and computation time scale well for \SpGEMMThreeD{} with $c=16, t=6$ when squaring \texttt{NaluR3} on Edison.
By contrast, communication time does not scale well for Sparse SUMMA (Figure~\ref{fig:scaleBarNaluR3}(a)), which eventually limits the scalability of 2D algorithms on higher concurrency.

\begin{figure}[!t]
   \centering
   \includegraphics[scale=.5]{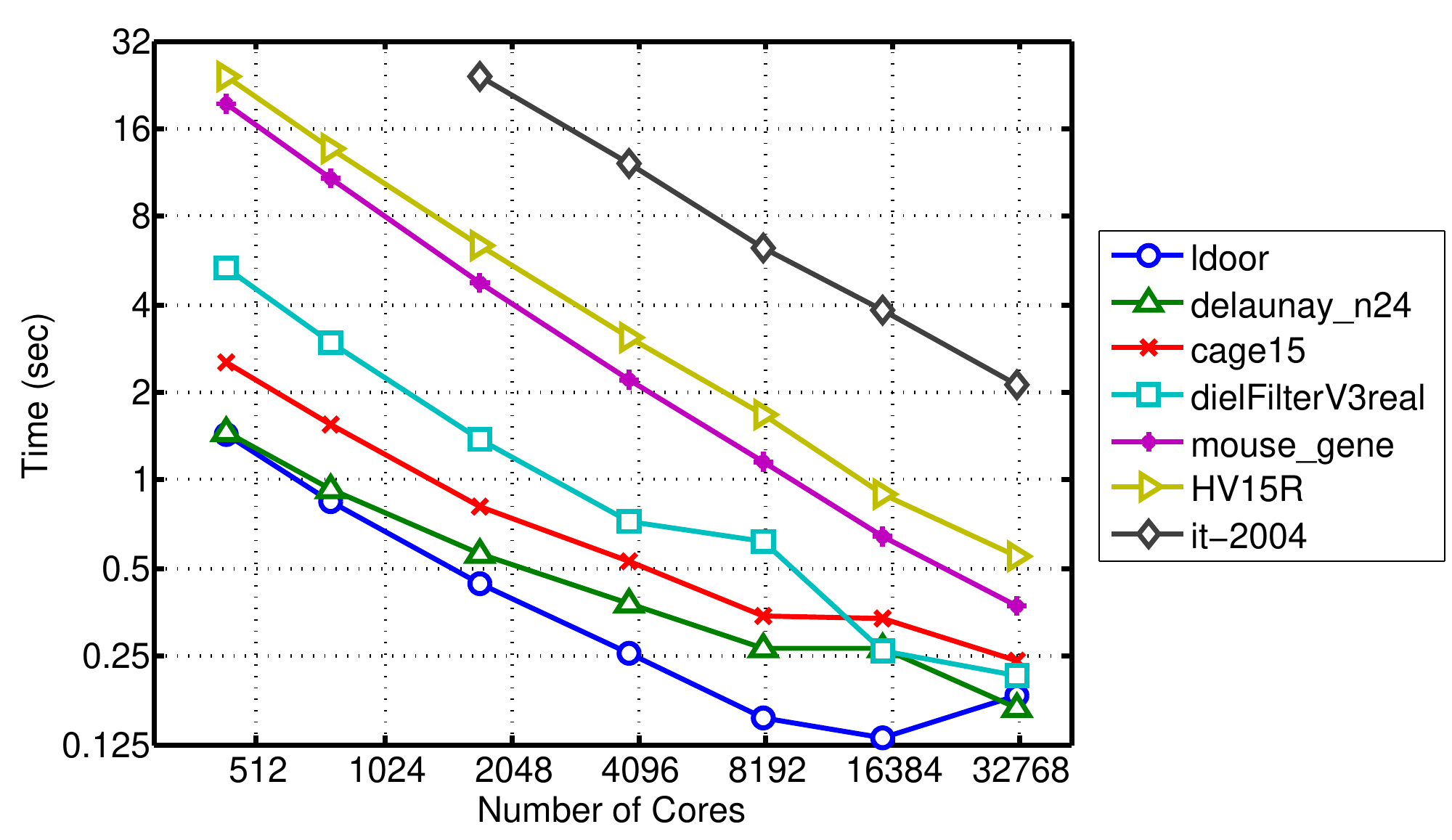} 
   \caption{Strong scaling of \SpGEMMThreeD{} with $c=16, t=6$ when squaring real matrices on Edison. 
   Large (e.g., \texttt{it-2004}) and dense (e.g., \texttt{mouse\_gene} and \texttt{HV15R}) matrices scale better than  small and sparse (e.g., \texttt{delaunay\_n24}) matrices. 
}
   \label{fig:realmats_scaling}
    \vsf
\end{figure}

\begin{figure}[!t]
   \centering
   \includegraphics[scale=.45]{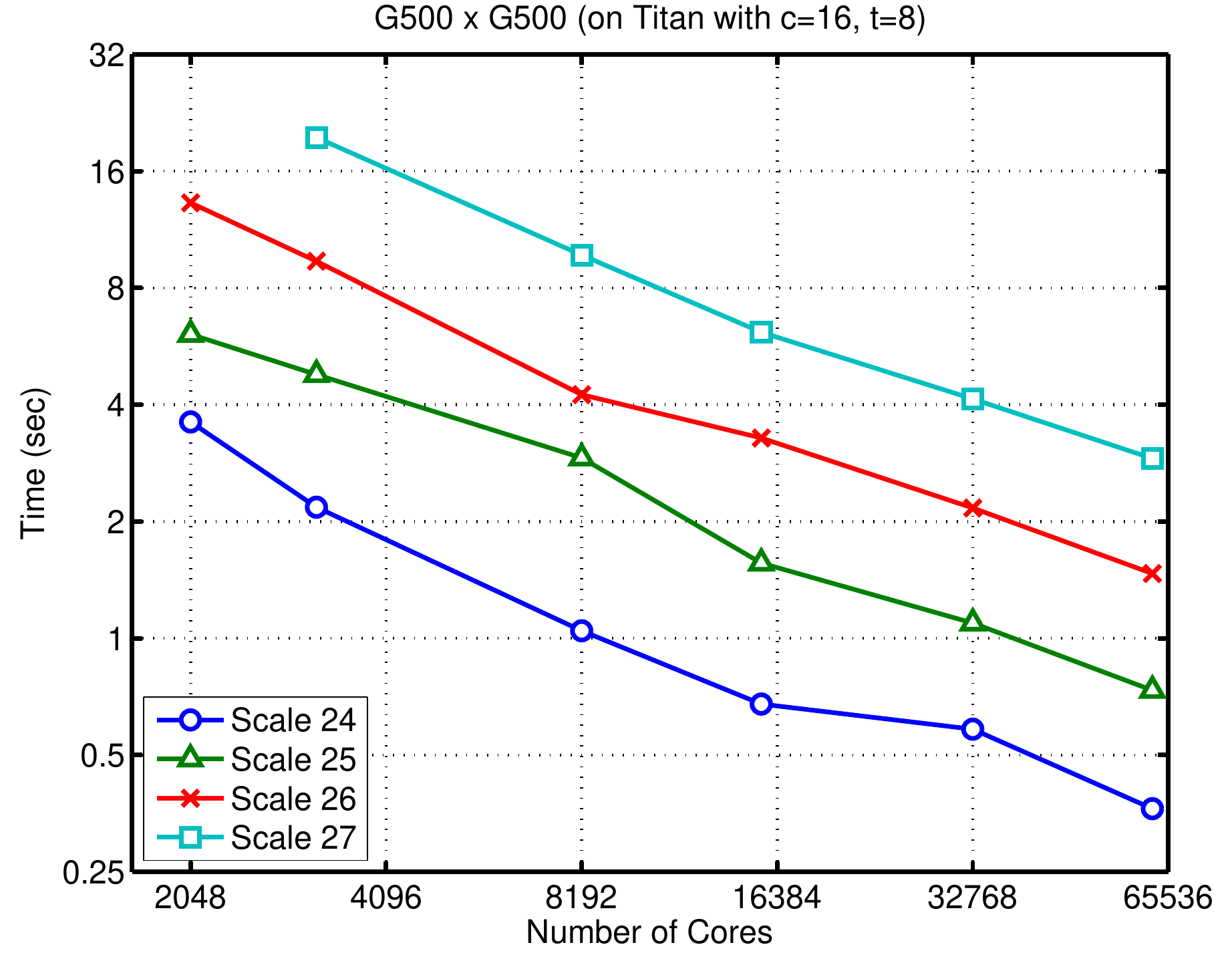} 
   \caption{Strong scaling of \SpGEMMThreeD{} with $c=16, t=8$ on Titan when multiplying two G500 matrices.}
   \label{fig:G500-strong-scaling-titan}
    \vsf
\end{figure}



\begin{figure}[!t]
   \centering
   \subfloat[][ER]{\includegraphics[scale=.35]{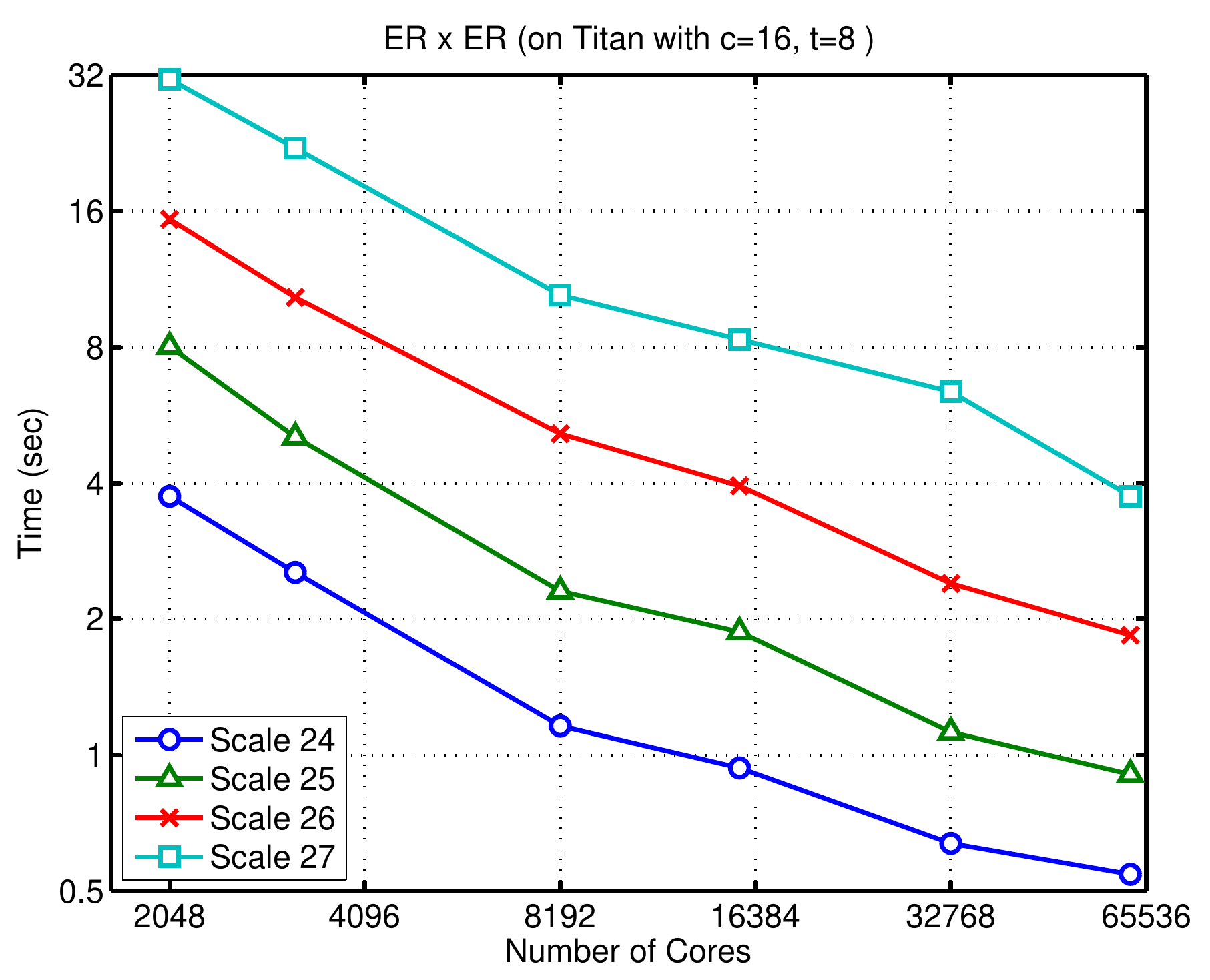} \label{fig:ER-strong-scaling-titan}}
   \subfloat[][SSCA]{\includegraphics[scale=.35]{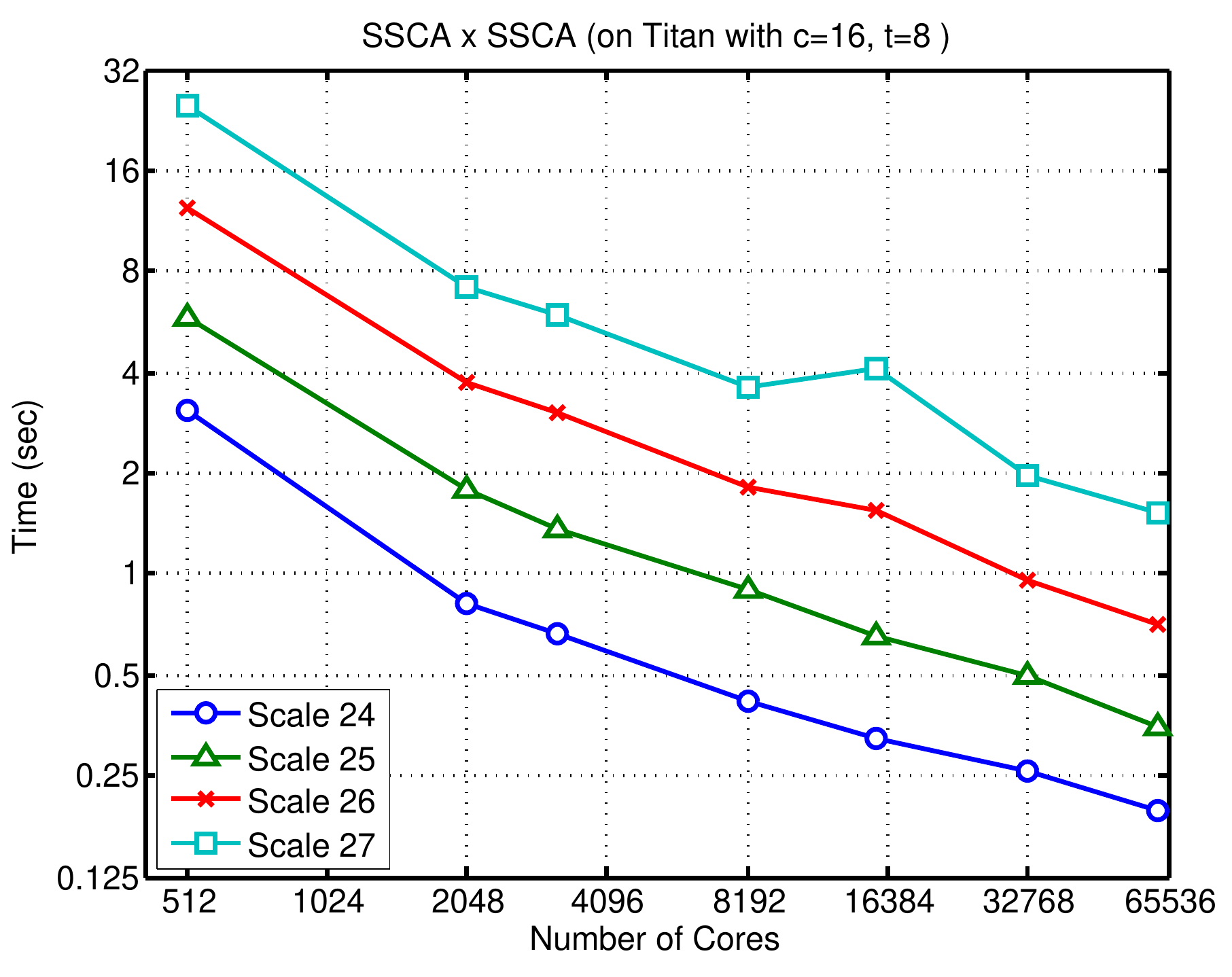} \label{fig:SSCA-strong-scaling-titan}} 
   \caption{Strong scaling of \SpGEMMThreeD{} with $c=16, t=8$ on Titan when multiplying two (a) ER and (b) SSCA matrices.}
   \vsf
\end{figure}

\subsubsection{Strong Scaling of \SpGEMMThreeD{}}
In this subsection, we show the strong scaling of \SpGEMMThreeD{} with the best parameters on higher concurrency ($c=16, t=6$ on Edison, and $c=16, t=8$ on Titan) when  multiplying real and random matrices. 
Figure~\ref{fig:realmats_scaling}  shows the strong scaling of \SpGEMMThreeD{} when squaring seven real matrices on Edison.
When we go from 512 to 32,768 cores ($64\times$ increase of cores), the average speedup of all matrices in Table~\ref{fig:testsuite} is about $27\times$ (min: $9\times$ for \texttt{delaunay\_n24}, max: $52\times$ for \texttt{mouse\_gene}, standard deviation: 16).
We observe that \SpGEMMThreeD{} scales better when multiplying larger (e.g., \texttt{it-2004}) and denser matrices (e.g., \texttt{mouse\_gene} and \texttt{HV15R}) because of the availability of more work. 
By contrast, \texttt{delaunay\_n24} is the sparsest matrix in Table~\ref{fig:testsuite} with $6$ nonzeros per column, and \SpGEMMThreeD{} does not scale well beyond 8,192 processor when squaring this matrix. 

Next, we discuss strong scaling of \SpGEMMThreeD{} for randomly generated square matrices whose dimensions range from $2^{24}$ to $2^{27}$. 
Figures \ref{fig:G500-strong-scaling-titan}, \ref{fig:ER-strong-scaling-titan}, and \ref{fig:SSCA-strong-scaling-titan} show the strong scaling of multiplying two structurally similar random matrices from classes G500, ER, and SSCA, respectively.
Once again, \SpGEMMThreeD{} scales better when multiplying larger (e.g., scale 27) and denser matrices (G500 and ER matrices have 16 nonzeros per row, but SSCA matrices have 8 nonzeros per row). Multiplying matrices with more nonzeros per row and column is expected to yield better scalability for these matrices.

\subsection{Multiplication with the Restriction Operator}
\label{sec:exp_restriction}
Multilevel methods are widely used in the solution of numerical and combinatorial problems.
Such methods construct smaller problems by successive coarsening. 
The simplest coarsening is graph contraction: a contraction step chooses two or more vertices in the original graph $G$ to become a single aggregate vertex in the contracted graph $G'$. 
The edges of $G$ that used to be incident to any of the vertices forming the 
aggregate become incident to the new aggregate vertex in $G'$.
 Constructing coarser representations in AMG or graph partitioning~\cite{brucegraph95} is a generalized graph contraction operation. 
This operation can be performed by multiplying the matrix representing the original fine domain (grid, graph, or hypergraph) by the restriction operator from the left and by the transpose of the restriction from the right~\cite{unifiedstarp}.

In our experiments, we construct the restriction matrix $\mathbf{R}$ using distance-2 maximal independent set computation, as described by Bell et al.~\cite{bell2012exposing}.
An independent set in a graph $G(V,E)$ is a subset of its vertices in which no two are neighbors. 
A maximal independent set (MIS) is an independent set that is not a subset of any other independent set.
MIS-2 is a generalization of MIS where no two vertices are distance-2 neighbors.  
In this scheme, each aggregate is defined by a vertex in MIS-2 and consists of the union of that vertex with its distance-1 neighbors. 

The linear algebraic formulation of Luby's randomized MIS algorithm~\cite{luby1986simple} was originally described earlier~\cite{lugowski2015parallel}. 
Here, we generalize it to distance-2 case, which is shown in Algorithm~\ref{alg:mis2} at a high level. \proc{MxV} signifies matrix-vector multiplication.
\proc{EwiseAdd} performs element-wise addition between two vectors, which amounts to a union operation among the index sets of those vectors. 
\proc{EwiseMult} is the element-wise
multiplication, which amounts to an intersection operation among the index sets.
For both \proc{EwiseAdd} and \proc{EwiseMult}, wherever
the index sets of two vectors overlap, the values for the overlapping indices are ``added" according to the binary function that is passed as the third parameter.
\Liref{sr1} finds the smallest random value among a vertex's neighbors using the semiring where scalar multiplication is overloaded with the operation that returns
the second operand and the scalar addition is overloaded with the minimum operation. \Liref{sr2} extends this to find the smallest random value among the 2-hop neighborhood.  
\Liref{newS} returns the new additions to MIS-2 if the random value of the second vector ($\mathsf{cands}$) is smaller.
\Liref{prune} removes those new additions, $\mathsf{newS}$, from the set of candidates. The rest of the computation is self-explanatory. 

\begin{algorithm}
\begin{algorithmic}[1]
\Require $\mA \in \mathbb{S}^{\dimN \times \dimN}, \mathsf{cands} \in \mathbb{S}^{1 \times \dimN}$
\Ensure $\mathsf{mis2} \in \mathbb{S}^{1 \times \dimN}$: distance-2 maximal independent set, empty in the beginning 
\Procedure{MIS2}{$\mA,  \mathsf{cands}, \mathsf{mis2}$}
\State $ \mathsf{cands} = 1:\dimN$ \Comment{all vertices are initially candidates} 
\While{$\Call{nnz}{\mathsf{cands}} >0 $} 
\State  $\Call{Apply}{\mathsf{cands}, \Call{Rand}}$ \Comment{generate random values} 
\State  $ \mathsf{minadj1} \gets \Call{MxV}{\mA, \mathsf{cands}, \Call{Semiring}{min, select2nd}} $ \lilabel{sr1}
\State  $ \mathsf{minadj2} \gets \Call{MxV}{\mA, \mathsf{minadj1}, \Call{Semiring}{min, select2nd}} $ \lilabel{sr2}
\State  $ \mathsf{minadj} \gets \Call{EwiseAdd}{\mathsf{minadj1}, \mathsf{minadj2}, \Call{Min}} $ \Comment{Union of minimums}
\State  $ \mathsf{newS} \gets \Call{EwiseMult}{\mathsf{minadj}, \mathsf{cands}, \Call{Is2ndSmaller}} $ \lilabel{newS} 
\State  $ \mathsf{cands} \gets \Call{EwiseMult}{\mathsf{cands}, \mathsf{newS}, \Call{select1st}} $  \lilabel{prune} 

\State  $ \mathsf{newS\_adj1} \gets \Call{MxV}{\mA, \mathsf{newS}, \Call{Semiring}{min, select2nd}} $ \lilabel{newsr1}
\State  $ \mathsf{newS\_adj2} \gets \Call{MxV}{\mA, \mathsf{newS\_adj1}, \Call{Semiring}{min, select2nd}} $ \lilabel{newsr2}
\State  $ \mathsf{newS\_adj} \gets \Call{EwiseAdd}{\mathsf{newS\_adj1}, \mathsf{newS\_adj2}, \Call{Any}} $ \Comment{Union of neighbors}
\State  $ \mathsf{cands} \gets \Call{EwiseMult}{\mathsf{cands}, \mathsf{newS\_adj}, \Call{select1st}} $  \lilabel{prune1} 
\State  $ \mathsf{mis2} \gets \Call{EwiseAdd}{\mathsf{mis2}, \mathsf{newS}, \Call{select1st}} $  \Comment{Add $\mathsf{newS}$ to $\mathsf{mis2}$}

\EndWhile
\EndProcedure
\end{algorithmic}
\caption{Pseudocode for MIS-2 computation in the language of matrices} \label{alg:mis2}
\end{algorithm}

Once the set $\mathsf{mis2}$ is computed, we construct the restriction matrix $\mathbf{R}$ by having each column
represent the union of a vertex in $\mathsf{mis2}$ with its distance-1 neighborhood. The neighborhood is calculated using another \proc{MxV} operation.
The remaining singletons are assigned to an aggregate randomly in order to ensure good load balance.
Consequently, $\mathbf{R}$ is of dimensions $\dimN \times \id{size}(\mathsf{mis2})$.

\begin{figure}[!t]
   \centering
   \includegraphics[scale=.55]{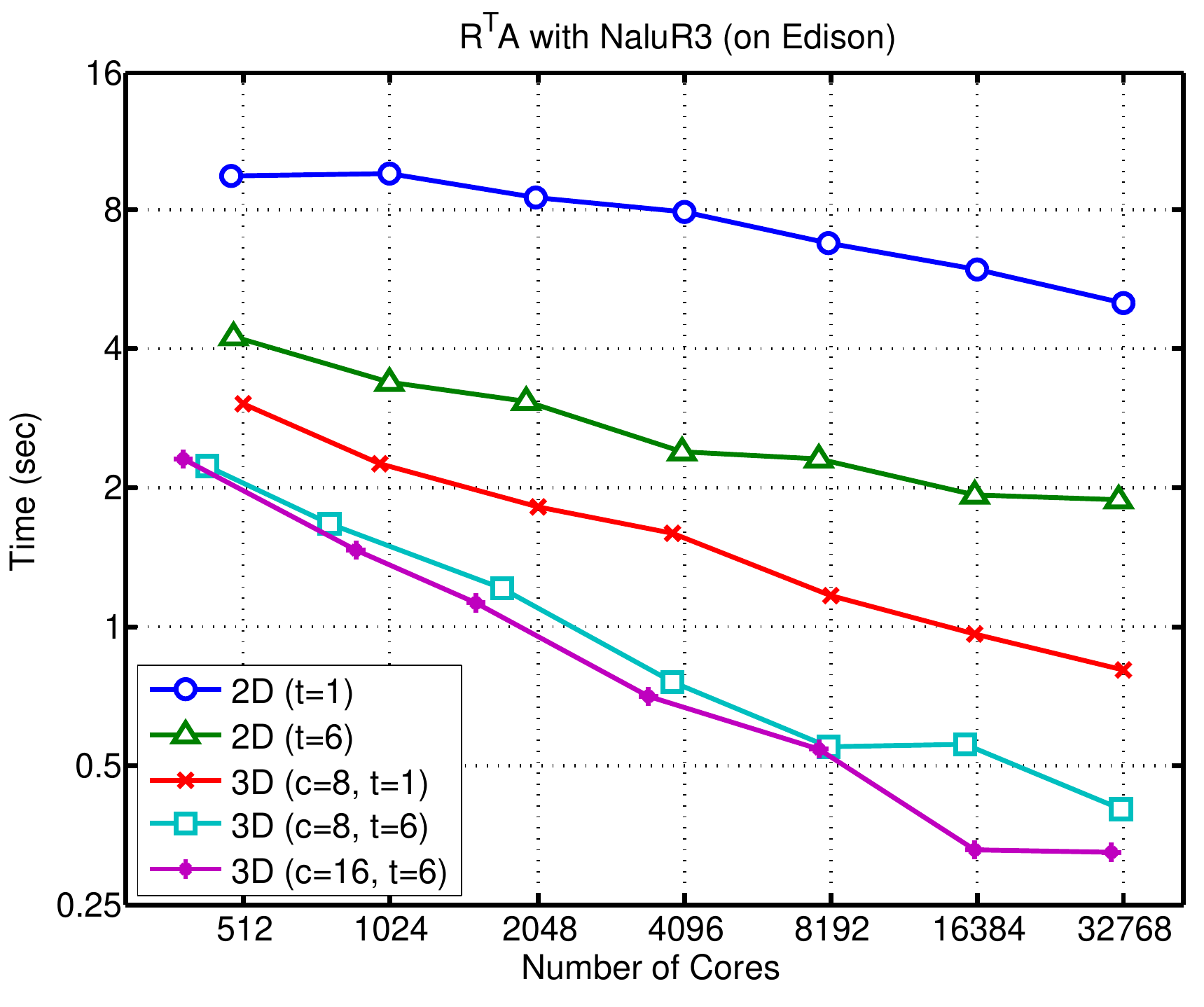} 
   \caption{Strong scaling of different variants of 2D and 3D algorithms  to compute $\mR\transpose\mA$ for \texttt{NaluR3} matrix on Edison.}
   \label{fig:rop_NaluR3_scaling}
    \vsf
\end{figure}

\begin{figure}[!t]
   \centering
   \subfloat[][nlpkkt160]{\includegraphics[scale=.31]{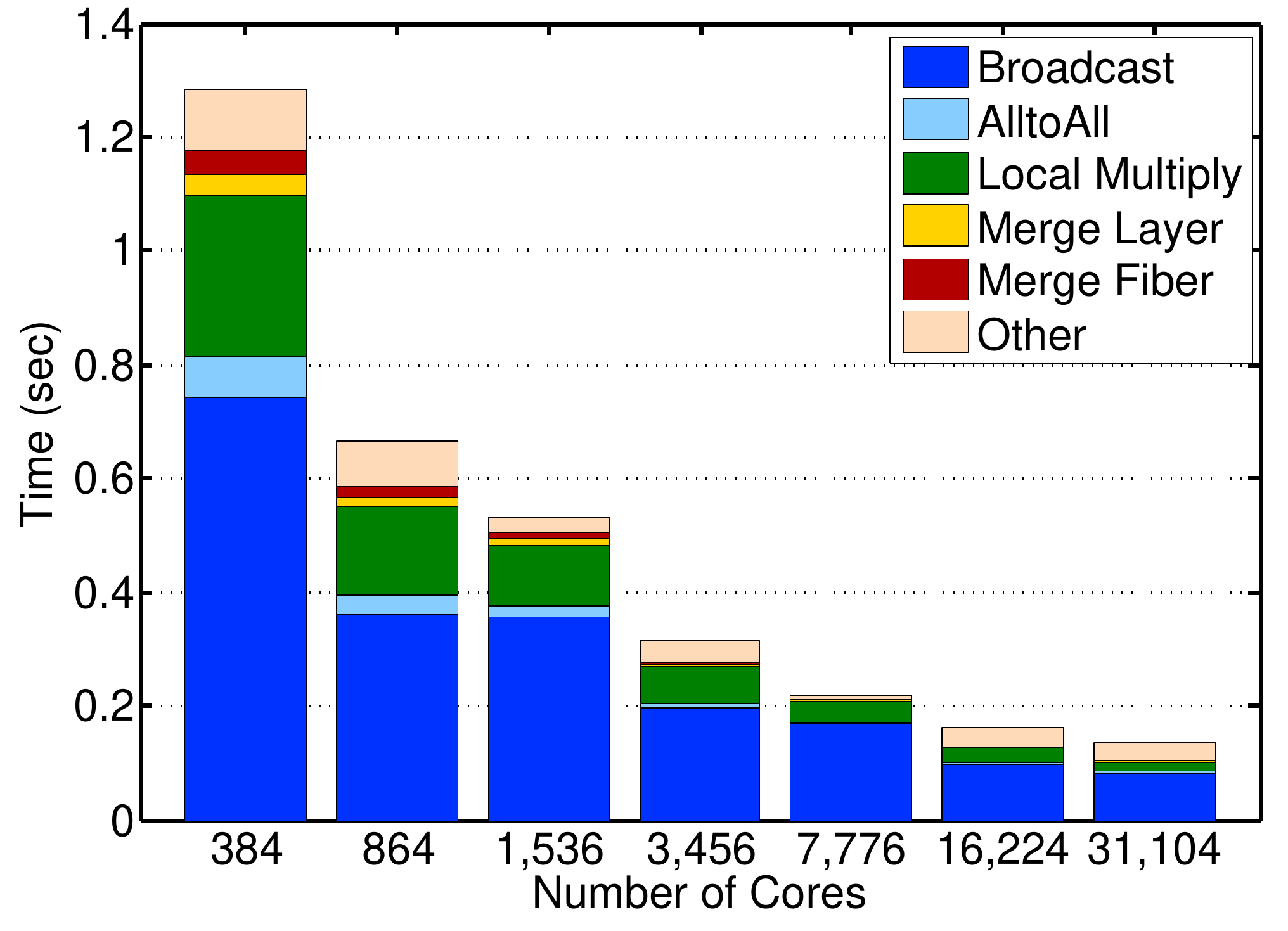}}
   \subfloat[][NaluR3]{\includegraphics[scale=.31]{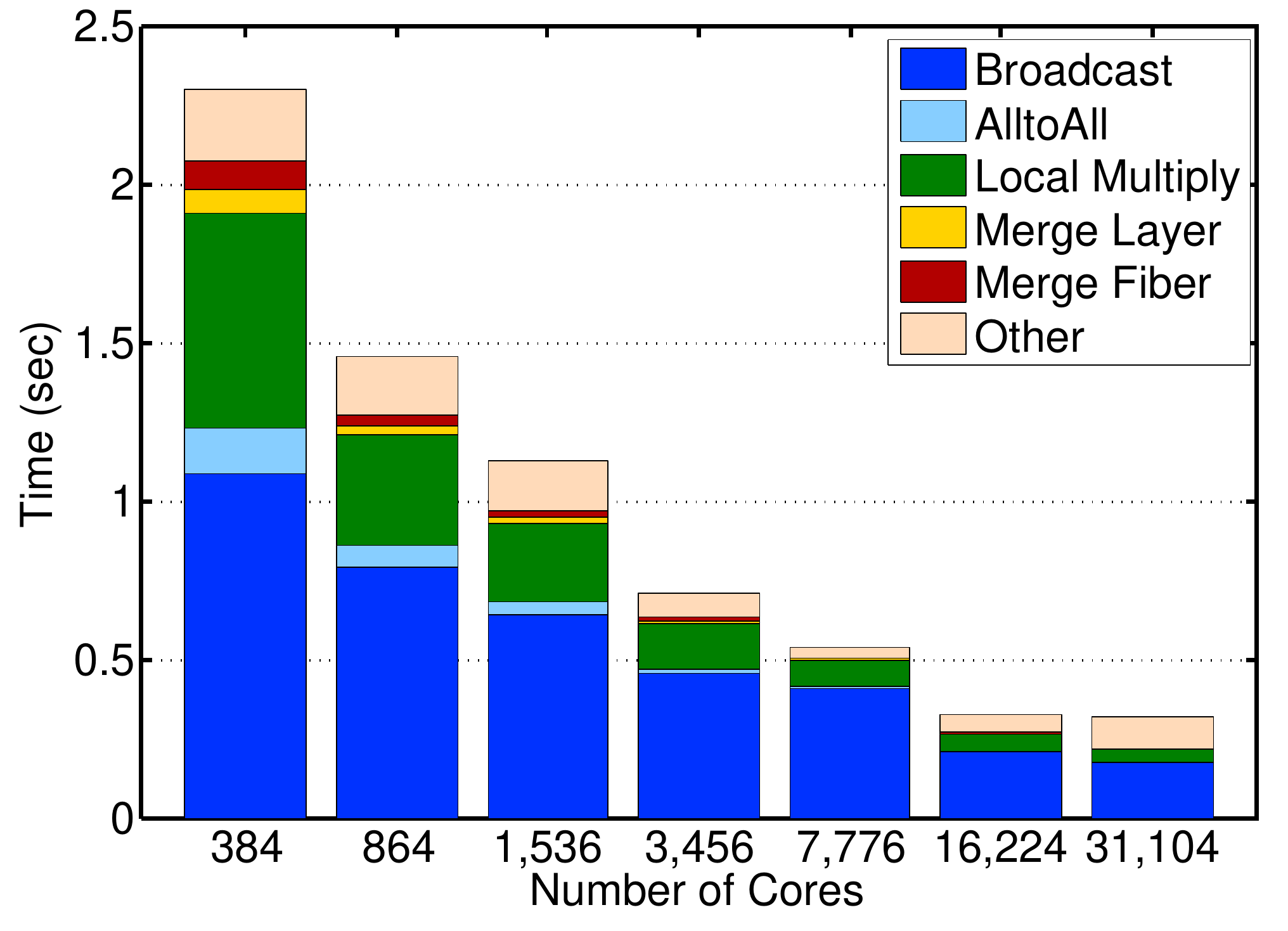}} 
   \caption{Breakdown of runtime spent by \SpGEMMThreeD{} to compute $\mR\transpose\mA$ for (a) \texttt{nlpkkt160}, and (b) \texttt{NaluR3} matrices with $c=16, t=6$ on Edison. Both communication and computation time scale well as we increase the number of cores.}
   \label{fig:rop_scaleBar}
    \vsf
\end{figure}

\begin{figure}[!t]
   \centering
   \includegraphics[scale=.55]{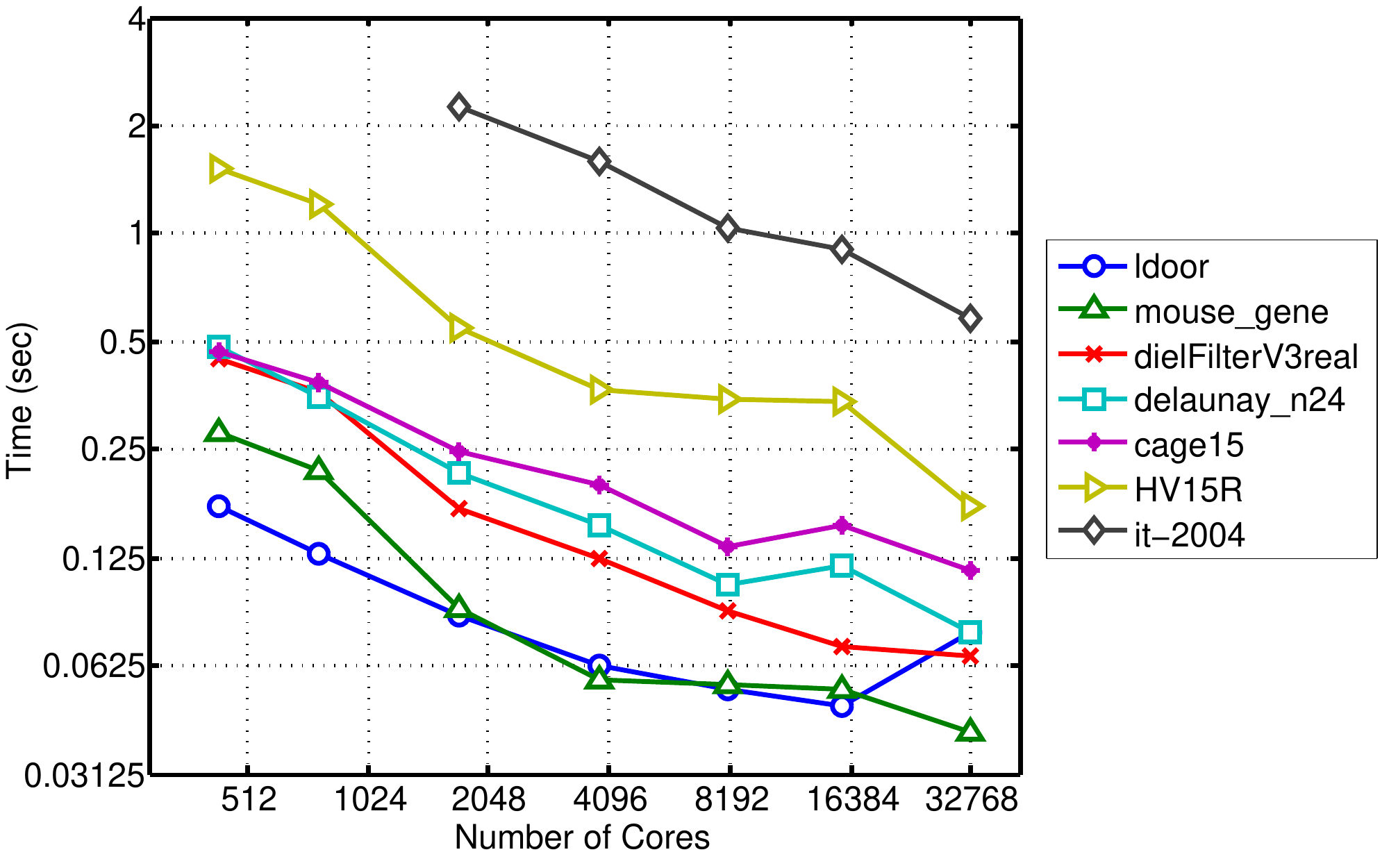} 
   \caption{Strong scaling of \SpGEMMThreeD{} to compute $\mR\transpose\mA$ with $c=16, t=6$ for seven real matrices on Edison. }
   \label{fig:rop_realmats_scaling}
    \vsf
\end{figure}

\subsubsection{Performance of Multiplying a Matrix with the Restriction Operator}
Figure~\ref{fig:rop_NaluR3_scaling} shows the strong scaling of different variants of 2D and 3D algorithms  when computing $\mR\transpose\mA$ with \texttt{NaluR3} matrix on Edison.
\SpGEMMThreeD{} with $c=16, t=6$ attains $7.5\times$ speedup  when we go from 512 cores to 32,768 cores, but other variants of 2D and 3D algorithms achieve lower speedups. 
Comparing the scaling of squaring \texttt{NaluR3} from Figure~\ref{fig:NaluR3_AA}, we observe moderate scalability of all variants of 2D and 3D algorithms when multiplying \texttt{NaluR3} with the restriction matrix. 
This is because the number of nonzeros in $\mR\transpose\mA$ is only 77 million, whereas $\dnnz(\mA^2)=2.1$ billion for \texttt{NaluR3} matrix (see Table~\ref{tab:stats}).
Hence, unlike squaring \texttt{NaluR3}, $\mR\transpose\mA$ computation does not have enough work to utilize thousands of cores.
However, the performance gap between 2D and 3D algorithms is larger when computing $\mR\transpose\mA$.
Figure~\ref{fig:rop_NaluR3_scaling} shows that \SpGEMMThreeD{} with $c=16, t=6$ runs $8\times$ and $16\times$ faster than non-threaded 2D algorithm on 512 and 32,768 cores, respectively.

Figure~\ref{fig:rop_scaleBar} shows the breakdown of runtime spent by \SpGEMMThreeD{} ($c=16, t=6$) to compute $\mR\transpose\mA$ for (a) \texttt{nlpkkt160}, and (b) \texttt{NaluR3} matrices on Edison.
We observe that when computing $\mR\transpose\mA$, \SpGEMMThreeD{} spends a small fraction of total runtime in the multiway merge routine. 
For example, on 384 cores in Figure~\ref{fig:rop_scaleBar}(b), 37\% of total time is spent on computation, and only about 7\% of total time is spent on multiway merge.
This is because $\dnnz(\mR\transpose\mA)$ is smaller than $nnz(\mA)$ for \texttt{NaluR3} matrix (also true for other matrices in Table~\ref{tab:stats}). 
Therefore, in computing $\mR\transpose\mA$, $\dnnz(\mR\transpose\mA)$ dominates the runtime of multiway merge while $\dnnz(\mA)$ dominates the local multiplication, making the former less computationally intensive. 
Hence, despite good scaling of local multiplication, the overall runtime is dominated by communication even on lower concurrency, thereby limiting the overall scaling on tens of thousands of cores.
By contrast, \SpGEMMThreeD{} spends 64\% of its total runtime in computation (with 40\% of the total runtime spent in multiway merge) when squaring \texttt{NaluR3} on 384 cores of Edison (Figure~\ref{fig:rop_scaleBar}(b)).
Hence, squaring of matrices shows better strong scaling than multiplying matrices with restriction operators.

Finally, Figure~\ref{fig:rop_realmats_scaling} shows the strong scaling of \SpGEMMThreeD{} ($c=16, t=6$) to compute $\mR\transpose\mA$ for other real matrices. 
\SpGEMMThreeD{} attains moderate speedups of up to $10\times$ when we got from 512 cores to 32,768 cores because of low computational intensity in the computation of $\mR\transpose\mA$.

\begin{figure}[!t]
   \centering
    \subfloat[][nlpkkt160]{\includegraphics[scale=.33]{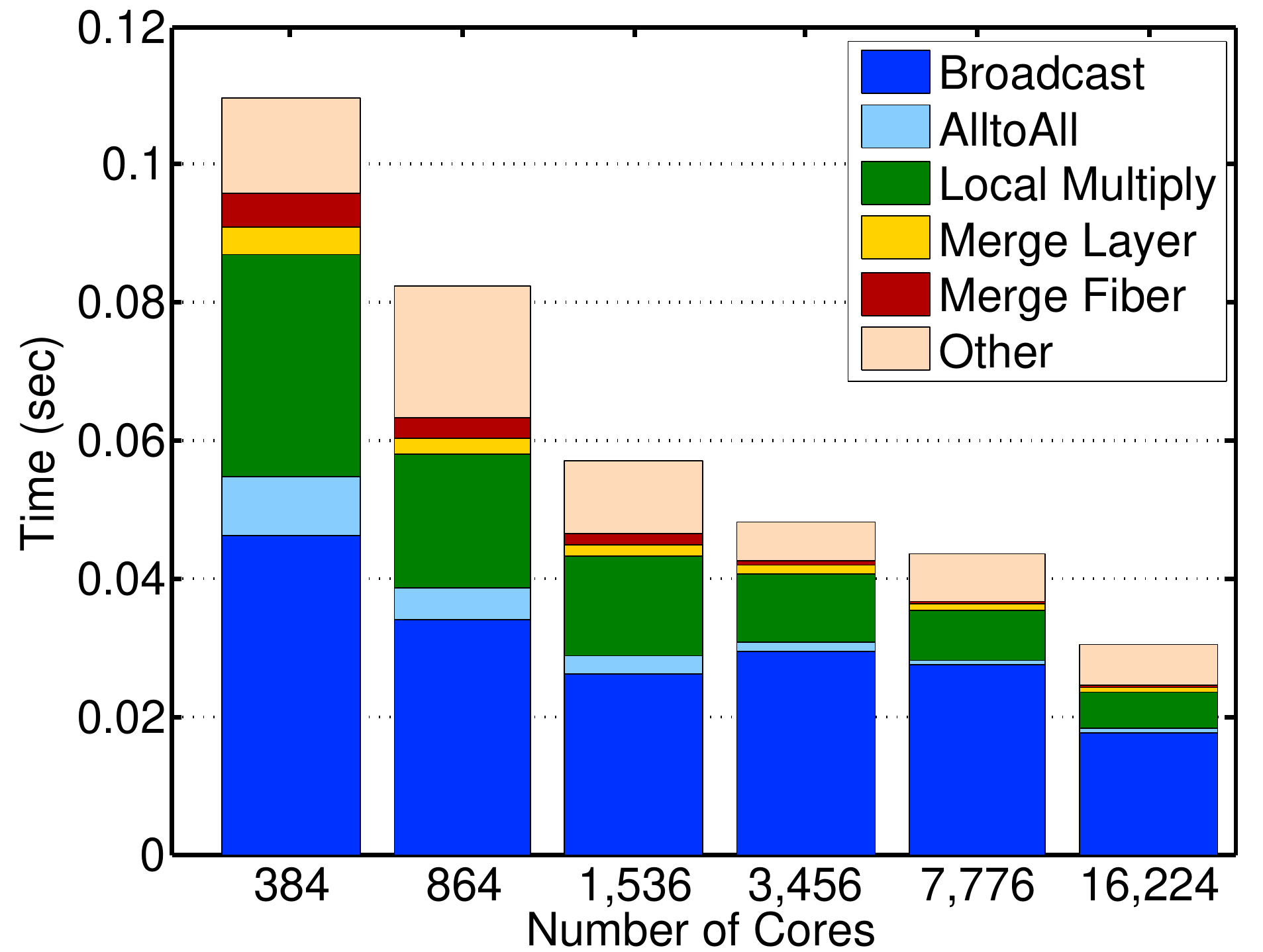}}
    ~~
   \subfloat[][NaluR3]{\includegraphics[scale=.33]{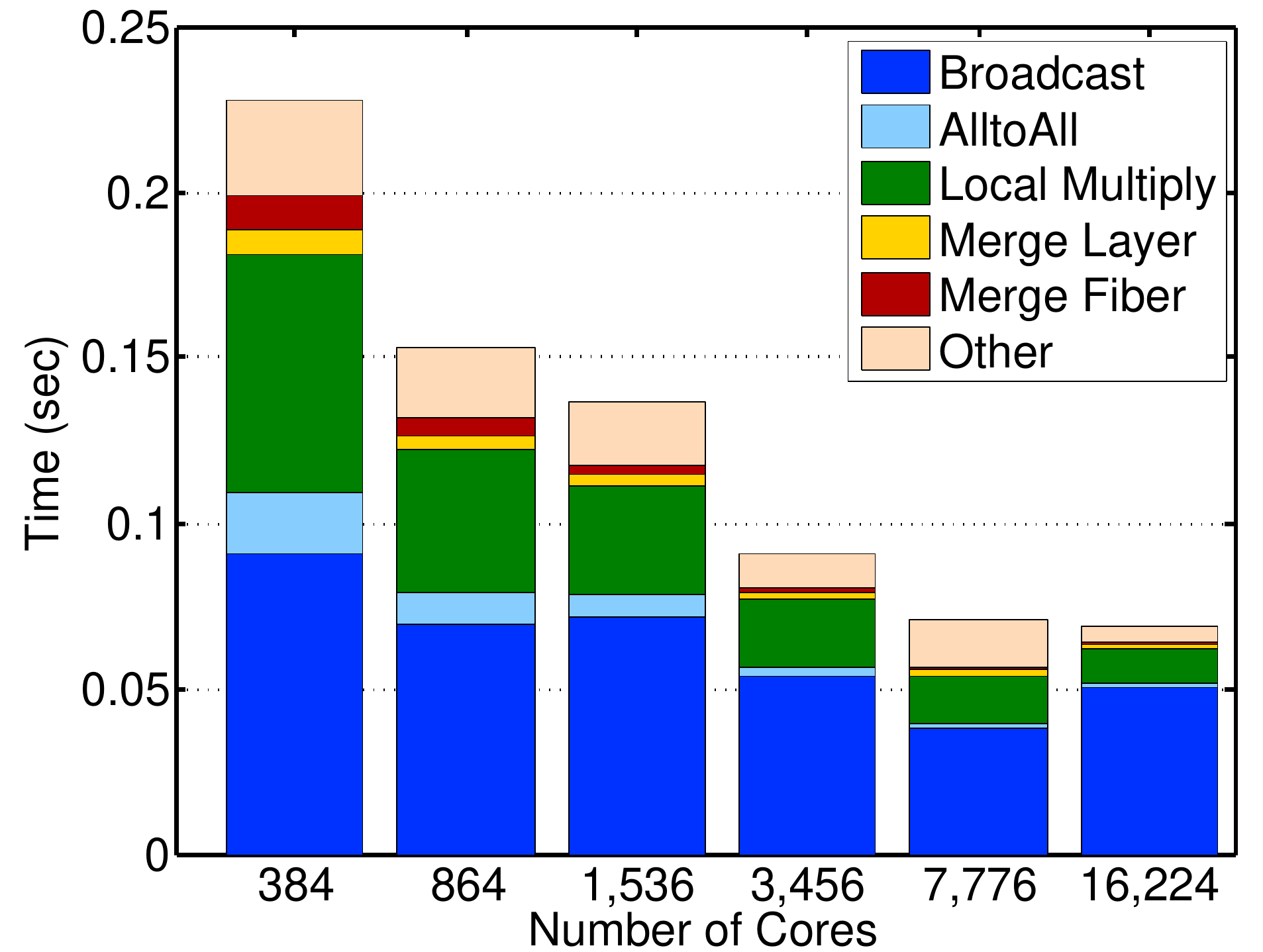}} 
   \caption{Breakdown of runtime spent by \SpGEMMThreeD{} to multiply $\mR\transpose\mA$ and $\mR$ for (a) \texttt{nlpkkt160}, and (b) \texttt{NaluR3} matrices with $c=16, t=6$ on Edison.}
   \label{fig:RAR_scaleBar}
    \vsf
\end{figure}

\subsubsection{Performance of Multiplying $\mR\transpose\mA$ and $\mR$}
Figure~\ref{fig:RAR_scaleBar} shows the scaling and breakdown of runtime spent by \SpGEMMThreeD{} ($c=16, t=6$) to multiply $\mR\transpose\mA$ and $\mR$ for (a) \texttt{nlpkkt160}, and (b) \texttt{NaluR3} matrices on Edison.
Even though $(\mR\transpose\mA)\mR$ computation can still obtain limited speedups on higher concurrency, the runtimes in Figure~\ref{fig:RAR_scaleBar} and the number of nonzeros in $\mR\transpose\mA\mR$ suggest that we might want to perform this multiplication on lower concurrency if necessary without degrading the overall performance.

 \begin{figure}[!t]
   \centering
   \subfloat[][nlpkkt160]{\includegraphics[scale=.42]{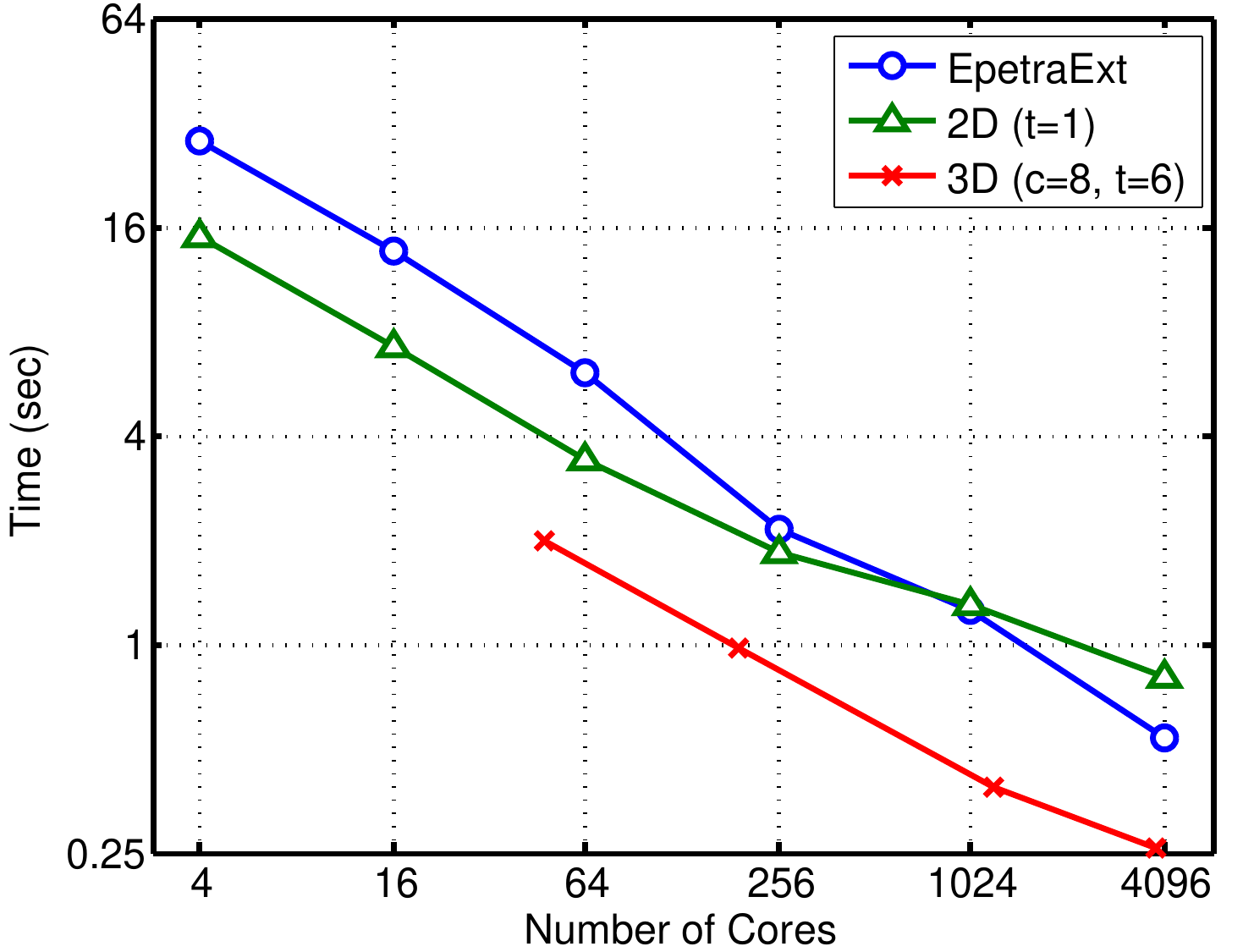}}
   ~~
   \subfloat[][NaluR3]{\includegraphics[scale=.42]{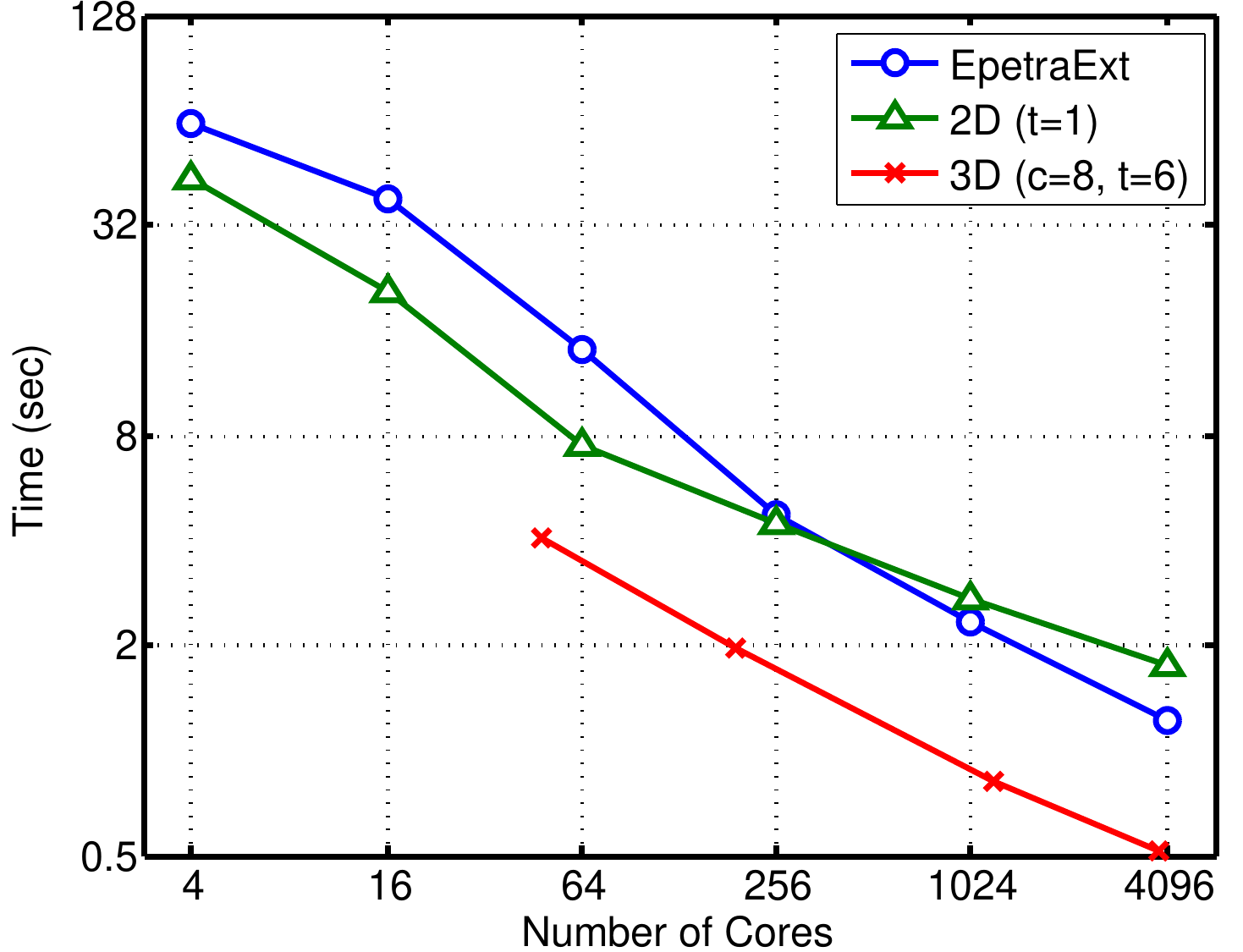}} 
   \caption{Comparison of Trilinos's EpetraExt package with 2D and 3D algorithms when computing $\mA\mR$ for \texttt{nlpkkt160} and \texttt{NaluR3} matrices on Edison.
   }
   \label{fig:trilinos_comp}
    \vsf
\end{figure}

\subsection{Comparison with Trilinos}
We compared the performance of our algorithms with the distributed-memory SpGEMM available in EpetraExt package of Trilinos.
We observed that SpGEMM in EpetraExt runs up to $3\times$ faster when we compute $\mA\mR$ instead of $\mR\transpose\mA$, especially on lower concurrency.
Hence, we only consider the runtime of $\mA\mR$ so that we compare against the best configuration of EpetraExt.
By contrast, our 2D and 3D algorithms are less sensitive to the order of matrix multiplication with less than $1.5\times$ performance improvement  in computing $\mA\mR$ over $\mR\transpose\mA$.
We use a random partitioning of rows to processors for EpetraExt runs.

Figure~\ref{fig:trilinos_comp} shows the strong scaling of EpetraExt's SpGEMM implementation and our 2D/3D algorithms when computing $\mA\mR$ on Edison.
On low concurrency, EpetraExt runs slower than the 2D algorithm, but the former eventually outperforms the latter on higher concurrency. 
However, on all concurrencies, the 3D algorithm with $c=8, t=6$ runs at least twice as fast as EpetraExt for these two matrices. 
We note that these matrices are structured with good separators where 1D decomposition used in EpetraExt usually performs better.
However, given the limitations of 1D decomposition for matrices without good separators, EpetraExt is not expected to perform well for graphs with power-law distributions~\cite{boman2013scalable}.
We have tried scale 24 Graph500 matrices in EpetraExt, but received segmentation fault in I/O.
We also tried other larger matrices, but EpetraExt could not finish reading the matrices from files in 24 hours, the maximum allocation limit for small jobs in Edison.
Hence, we compare with EpetraExt on problems that it excels (AMG style reduction with matrices having good separators) and even there our 3D algorithm does comparably better.
We could have separated the diagonal for better scaling  performance~\cite{gemmexp}, but we decided not to as if would break the ``black box" nature of our algorithm.

\section{Conclusions and Future Work}
We presented the first implementation of the 3D parallel formulation of sparse matrix-matrix multiplication (SpGEMM).
Our implementation exploits inter-node parallelism within a third processor grid dimension as well as thread-level parallelism within the node. 
It achieves higher performance compared to other available formulations of distributed-memory SpGEMM, without compromising 
flexibility in the numbers of processors that can be utilized. In particular, by varying the third processor dimension 
as well as the number of threads, one can run our algorithm on many 
processor counts.

The percentage of time spent in communication (data movement) is significantly lower in our new implementation compared to a 2D implementation.
This is advantageous for multiple reasons. First, the bandwidth for data movement is expected to increase at a slower rate than other system components, 
providing a future bottleneck. Second, communication costs more energy than computation~\cite[Figure 5]{kogge2013exascale}. 
Lastly, communication can be hidden by overlapping it with local computation, up to the time it takes to do the local computation. For example, up to
$100\%$ performance increase can be realized with overlapping if the communication costs 50\% of overall time. However, if the communication costs 80\%
of the time, then overlapping can only increase performance by up to $25\%$. Overlapping communication with computation as well as exploiting task-based
programming models are subject to future work.

Our 3D implementation inherits many desirable properties of the 2D matrix decomposition, such as resiliency against matrices with skewed degree distribution that
are known to be very challenging for traditional 1D distributions and algorithms. However, the 3D formulation also avoids some of the pitfalls of 2D algorithms, such as their relatively
poor performance on structured matrices (due to load imbalance that occurs on the processor on the diagonal), by exploiting parallelism
along the third dimension. This enabled our algorithm to beat a highly-tuned 1D implementation (the new EpetraExt) on structured matrices, without
resorting to techniques such as matrix splitting that were previously required of the 2D algorithm for mitigating the aforementioned load imbalance~\cite{gemmexp}. 

Our experimental results indicate that at large concurrencies, performance of the inter-node communication collectives becomes the determinant factor in overall performance. Even though work on the 
scaling of collectives on subcommunicators is under way, we believe that the effect of simultaneous communication on several subcommunicators are not well studied and should be the
subject of further research.
 
\section*{Acknowledgments}
We sincerely thank the anonymous reviewers whose feedback greatly improved the presentation and clarity of this paper.

This material is based upon work supported by the U.S. Department of Energy, Office of Science, Office of Advanced Scientific Computing Research, Applied Mathematics program under contract number DE-AC02-05CH11231.

This research was supported in part by an appointment to the Sandia National Laboratories Truman Fellowship in National Security Science and Engineering, sponsored by Sandia Corporation (a wholly owned subsidiary of Lockheed Martin Corporation) as Operator of Sandia National Laboratories under its U.S. Department of Energy Contract No. DE-AC04-94AL85000.

The research of some of the authors was supported by the U.S. Department of Energy Office of Science, 
Office of Advanced Scientific Computing Research, Applied Mathematics program under award DE-SC0010200, by the 
U.S. Department of Energy Office of Science, 
Office of Advanced Scientific Computing Research, X-Stack program under awards 
DE-SC0008699, DE-SC0008700, and AC02-05CH11231, and by 
DARPA award HR0011-12-2-0016, with contributions from Intel, Oracle, and MathWorks.

Research is supported by grants 1878/14, and 1901/14 from the Israel Science Foundation (founded by the Israel Academy of Sciences and Humanities) and grant 3-10891 from the Ministry of Science and Technology, Israel. Research is also supported by the Einstein Foundation and the Minerva Foundation. This work was supported by the HUJI Cyber Security Research Center in conjunction with the Israel National Cyber Bureau in the Prime Minister's Office. This paper is supported by the Intel Collaborative Research Institute for Computational Intelligence (ICRI-CI). This research was supported by a grant from the United States-Israel Binational Science Foundation (BSF), Jerusalem, Israel.
 
This research used resources of the National Energy Research Scientific Computing Center, which is supported by the Office of Science of the U.S. Department of Energy under Contract No. DE-AC02-05CH11231, and resources of the Oak Ridge Leadership Facility at the Oak Ridge National Laboratory, which is supported by the Office of Science of the U.S. Department of Energy under Contract No. DE-AC05-00OR22725.

\bibliographystyle{plain}
\bibliography{M104253}
\end{document}